\documentclass[twocolumn,pra,amssymb,showpacs]{revtex4-1}
\usepackage{amsmath}
\usepackage{graphicx}
\usepackage{subfigure}

\def\a{\mathbf{a}}
\def\b{\mathbf{b}}

\begin{document}

\title{Nonlocality and entanglement in qubit systems}

\author{J. Batle and M. Casas}

\email{E-mail address (JB): vdfsjbv4@uib.es}
\affiliation{Departament de F\'{\i}sica and IFISC-CSIC, Universitat de les Illes Balears, 07122 Palma de Mallorca, Spain}

\date{\today}

\begin{abstract}

Nonlocality and quantum entanglement constitute two special aspects of the quantum correlations existing in quantum systems, 
which are of paramount importance in quantum-information theory. Traditionally, they have been regarded as identical 
(equivalent, in fact, for pure two qubit states, that is, {\it Gisin's Theorem}), yet they constitute different resources. 
Describing nonlocality by means of the violation of several Bell inequalities, we obtain by direct optimization those states of two qubits 
that maximally violate a Bell inequality, in terms of their degree of mixture as measured by either their participation
ratio $R=1/Tr(\rho^2)$ or their maximum eigenvalue $\lambda_{max}$. This optimum value is obtained as well, which coincides 
with previous results. Comparison with entanglement is performed too. An example of an application is given in the XY model. 
In this novel approximation, we also concentrate on the nonlocality for linear combinations of pure states of two qubits, providing a closed
form for their maximal nonlocality measure. The case of Bell diagonal mixed states of two qubits is also extensively studied. 
Special attention concerning the connection between nonlocality and entanglement for mixed states of two qubits is paid to the 
so called maximally entangled mixed states. Additional aspects for the case of two qubits are also described in detail.
Since we deal with qubit systems, we will perform an analogous study for three qubits, employing similar tools. 
Relation between distillability and nonlocality is explored quantitatively for the whole space of states of three qubits. 
We finally extend our analysis to four qubit systems, where nonlocality for generalized Greenberger-Horne-Zeilinger states of 
arbitrary number of parties is computed.

\end{abstract}
\maketitle

\section{INTRODUCTION}

Schr\"odinger's \cite{Schro} modern notion of entangled state historically appeared within the debate of the 
paradox posed by Einstein, Podolsky and Rosen (EPR) \cite{EPR}. EPR pointed out
the possible lack of completeness of the newborn theory of quantum mechanics. In their famous paper
they suggested a description of nature (called ``local realism'') which assigns an independent
and objective reality to the properties of separated parties of a composite physical system.
EPR applied the criterion of local realism to predictions associated with an entangled state,
a state that cannot be described solely in terms of the properties of its subsystems, to conclude
that quantum mechanics was incomplete. Schr\"odinger, instead, regarded entanglement as 
{\it the} characteristic feature of quantum mechanics. Quantitatively though,
no measure for this ``quantum strangeness'' was provided at the time.

The most significant progress toward the resolution of the EPR debate came with Bell's work. 
Bell \cite{Bell1,Bell2} proved the impossibility of reproducing all correlations observed
in composite quantum systems using models similar to that of EPR. In fact,
Bell showed that local realism, in the form of local variable models (LVM), implies
constraints on the predictions of spin correlations in the form of inequalities, also known
as Bell's inequalities, which can be violated by quantum mechanics. That is why quantum mechanics 
is regarded as being inherently nonlocal.

If quantum mechanics could be described by LVM, then the correlation values measured 
between parties could be reproduced assuming the corresponding operators had already a definite 
value previous to measurement. Let us consider the case of the two-party two-outcome scenario. 
Two distant observers possessing a distributed state $\rho_{AB}$ exploit the correlations arising from

\begin{equation}\label{BellQM}
P(a,b|A,B;\rho_{AB})= Tr[\rho_{AB}(\Pi^{A}_{a} \otimes \Pi^{B}_{b})]
\end{equation}

\noindent quantum mechanically. $\Pi^{A}_{a}$ and $\Pi^{B}_{b}$ are positive operators
involving measurements of observables $A$ and $B$, with outcomes $a$ and $b$ satisfying
$A\Pi^{A}_{a}=a\Pi^{A}_{a}$ and $B\Pi^{B}_{b}=b\Pi^{B}_{b}$. If a LVM could mimic the same
correlations

\begin{equation}\label{BellClass}
P(a,b|A,B;\rho_{AB})= \sum_{\lambda} P(a|A,\lambda) P(b|B,\lambda) \mu(\lambda),
\end{equation}

\noindent with $\mu(\lambda)$ being a probability measure for the classical variable
$\lambda$ and $P(a|A,\lambda)$ and $P(b|B,\lambda)$ local functions, therefore the use of
state $\rho_{AB}$ would provide no improvement over classical resources. Thus it was clear
that the notion of nonlocality of a state $\rho_{AB}$ would emerge if there existed no
LVM that through (\ref{BellClass}) could reproduce the quantum mechanical results 
of (\ref{BellQM}). Nonlocality was a character of entangled states via violation of a Bell inequality.

Ever since Bell's contribution, entanglement and nonlocality became similar terms. The
nonlocal character of entangled states was clear for pure states. In fact, all entangled pure
states of two qubits violate the CHSH inequality and therefore are nonlocal \cite{Gisin}. The situation
became more involved when Werner \cite{Werner} discovered that while entanglement is necessary for
a state to be nonlocal, for mixed states is not sufficient. He also introduced the usual (modern)
definition of entangled state: given two parties A and B, a shared state $\rho_{AB}$ is
termed unentangled or separable if it cannot be expressed as the mixture of product states

\begin{equation}\label{rhosep}
\rho_{AB}=\sum_{i=1}^{N} p_i |\psi_{A}^{i}\rangle \langle \psi_{A}^{i}| \otimes |\psi_{B}^{i}\rangle \langle \psi_{B}^{i}|,
\end{equation}

\noindent that is, when its preparation does not require a nonlocal quantum resource. This definition, 
in spite of its clear physical meaning, is somewhat
impractical, since tests to distinguish separable from entangled
states are complicated. 

With the advent of quantum-information theory (QIT), the interest in
entanglement has dramatically increased over the years since it lies at the basis of some of the
most important processes and applications studied by QIT such as quantum cryptographic key distribution \cite{E91},
quantum teleportation \cite{BBCJPW93}, superdense coding \cite{BW93} and quantum computation \cite{EJ96,BDMT98},
among many others which possess no classical counterpart. All these tasks require distributed
quantum correlations between parties, and the only means available in nature are
entangled states. Using a modern nomenclature, when a quantum state cannot be prepared using only
local operations and classical communications (LOCC), it is said that it possesses quantum correlations
and the state is entangled. Spatially separated observers sharing an entangled state and performing
measurements on them may induce nonlocal correlations which cannot be simulated by local means
(violate Bell inequalities).

Confusion between nonlocality and entanglement appeared during the finer study of the usefulness of
quantum correlations. Entanglement is commonly viewed as a useful resource
for various information-processing tasks. Yet, there exist certain tasks, such as device-independent
quantum key distribution \cite{device} and quantum communication complexity problems \cite{comcomplex}, which can only be carried
out provided the corresponding entangled states exhibit nonlocal correlations.
Then we are naturally led to the question of whether nonlocality and entanglement constitute two different
resources.

The purpose of the present work is to shed some light upon the relation between entanglement
and nonlocality through the maximal violation of a Bell inequality for two, three and four qubit systems. 
Throughout the article, and in order to avoid confusion, we will refer to the
quantity ``nonlocality'' as being equivalent to ``maximum violation of a Bell
inequality''. However, the usual meaning of nonlocality (or that of a
nonlocal state) as a concept involving the mere violation of a Bell inequality remains the same. 
Although detection and characterization of entanglement is far from being complete, its status is more 
developed than that of nonlocality. The problems experienced in defining a unique 
measure of entanglement in the multipartite case, in the form of partitions in the system, disappear in the 
nonlocality case since there exists well defined Bell inequalities for multipartite systems of arbitrary number of qubits. 
This fact makes the study of nonlocality conceptually and quantitatively easier.

This paper is organized as follows. In Section II we review recent results concerning nonlocality
in bipartite physical systems and concentrate on the CHSH Bell inequality \cite{CHSH} for two qubits.
We also obtain, after a direct optimization over the observers' settings, the family of two qubit mixed 
states that optimizes the violation of the CHSH inequality for a given degree of mixture, as well as the 
concomitant optimal value for CHSH. This novel approach recovers and extends previous results found in the literature.	 
Our results concerning the connection between nonlocality and entanglement for mixed states of two qubits also pay special 
attention to the case of the so called maximally entangled mixed states (MEMS). 
As far as the duality entanglement-nonlocality is concerned, these very same results find too 
interesting echoes in a well known condensed matter system, namely, the infinite $XY$ model.
How nonlocality can be present in linear combinations of pure states of two qubits is
also reported, where the optimal value of the violation of the CHSH inequality is obtained for any
given superposition. Parallelism with the role of entanglement of superposition is also discussed. 
Section III is devoted to the study of entanglement and nonlocality for three qubit systems, employing similar tools. 
Relation between distillability and nonlocality is explored quantitatively for the whole space of states of three qubits.
Section IV extends the present subject of study to four qubit systems, where nonlocality for 
generalized Greenberger-Horne-Zeilinger (GHZ) states of arbitrary number of parties is computed. 
Finally, some conclusions are drawn in Section V.

\section{TWO QUBITS}

\subsection{Nonlocality and the CHSH Bell inequality}


LVM cannot exhibit arbitrary correlations. Mathematically, the conditions these correlations
must obey can always be written as inequalities --the Bell inequalities-- satisfied for the joint
probabilities of outcomes. We say that a quantum state $\rho$ is nonlocal if and only if there
are measurements on $\rho$ that produce a correlation that violates a Bell inequality.

Most of our knowledge on Bell inequalities and their quantum mechanical violation is based
on the CHSH inequality \cite{CHSH}. With two dichotomic observables per party, it is the
simplest \cite{Collins} (up to local symmetries) nontrivial Bell inequality for the bipartite case with
binary inputs and outcomes. Let $A_1$ and $A_2$ be two possible measurements on A side whose
outcomes are $a_j\in \lbrace -1,+1\rbrace$, and similarly for the B side. Mathematically, it can
be shown that, following LVM (\ref{BellClass}),
$|{\cal B}_{CHSH}^{LVM}(\lambda)|=|a_1b_1+a_1b_2+a_2b_1-a_2b_2|\leq 2$. Since $a_1$($b_1$)
and $a_2$($b_2$) cannot be measured simultaneously, instead one estimates after randomly
chosen measurements the average value ${\cal B}_{CHSH}^{LVM} \equiv \sum_{\lambda} {\cal B}_{CHSH}^{LVM}(\lambda) \mu(\lambda)=
E(A_1,B_1)+E(A_1,B_2)+E(A_2,B_1)-E(A_2,B_2)$, where $E(\cdot)$ represents the expectation value.
Therefore the CHSH inequality reduces to

\begin{equation} \label{CHSH_LVM}
|{\cal B}_{CHSH}^{LVM}| \leq 2.
\end{equation}

Quantum mechanically, since we are dealing with qubits, these observables reduce
to ${\bf A_j}({\bf B_j})={\bf a_j}({\bf b_j}) \cdot {\bf \sigma}$, where ${\bf a_j}({\bf b_j})$
are unit vectors in $\mathbb{R}^3$ and ${\bf \sigma}=(\sigma_x,\sigma_y,\sigma_z)$ are the usual
Pauli matrices. Therefore the quantal prediction for (\ref{CHSH_LVM}) reduces to the expectation
value of the operator ${\cal B}_{CHSH}$

\begin{equation} \label{CHSH_QM}
{\bf A_1}\otimes {\bf B_1} + {\bf A_1}\otimes {\bf B_2} 
+ {\bf A_2}\otimes {\bf B_1}  -  {\bf A_2}\otimes {\bf B_2}.
\end{equation}

\noindent Tsirelson showed \cite{Tsirelson} that CHSH inequality (\ref{CHSH_LVM}) is 
maximally violated by a multiplicative
factor $\sqrt{2}$ (Tsirelson's bound) on the basis of quantum mechanics. In fact, it is
true that $|Tr(\rho_{AB}{\cal B}_{CHSH})|\leq 2\sqrt{2}$ for all observables ${\bf A_1}$,
${\bf A_2}$, ${\bf B_1}$, ${\bf B_2}$, and all states $\rho_{AB}$. Increasing the
size of Hilbert spaces on either A and B sides would not give any advantage in the
violation of the CHSH inequalities. In general, it is not known how to calculate the best
such bound for an arbitrary Bell inequality, although several techniques have
been developed \cite{Toner}.

A good witness of useful correlations is, in many cases, the violation of a Bell inequality
by a quantum state. But not all entangled states are nonlocal. Although this is
the case for pure states of two qubits (CHSH inequality violation), Werner showed that
it cannot be generalized to mixed states. After introducing the states which are now
called Werner states

\begin{equation} \label{Werner}
\rho_{W}=p |\psi^{-}\rangle \langle \psi^{-}| + (1-p)\frac{I}{4},
\end{equation}

\noindent where $|\psi^{-}\rangle$ is the singlet state and $I$ is the $4\times 4$ identity,
he provided a LVM for measurement outcomes for some entangled states of this
family. Although promising new results have been obtained recently \cite{Acin,Vertesi}, even in the
simplest case of Werner states of two qubits (\ref{Werner}), it is in general extremely difficult to
determine whether an entangled state has a LVM or not, since finding all Bell inequalities
is a computationally hard problem \cite{hard1,hard2}.


Therefore we shall consider the optimization of the violation of the CHSH inequality over
the observer's settings as a definitive measure for both signaling
and quantifying nonlocality in two qubit systems.

\subsection{Maximal violation of the CHSH inequality and mixedness for two qubits}


What is the maximum violation of $B_{CHSH} (=Tr(\rho{\cal B}_{CHSH}) \leq 2$) for a given state $\rho$?\newline 

\noindent Before any attempt to proceed with a definite optimization program, we shall undertake a detailed analysis of the special form for 
$B_{CHSH}=Tr(\rho{\cal B}_{CHSH})$, the basic quantity we are about to deal with. 
The nature of any bipartite mixed state of two qubits is described by a positive, semi definite 
matrix, whose eigenvalues $\{ \lambda_i \}$ are such that they $0\leq \lambda_i \leq 1$ and $\sum_i \lambda_i =1$. 
In other words, the complete description of the density matrix $\rho$ necessitates $4^2-1=15$ real parameters. 

Usually, the preferred basis for two qubit states is the so called computational basis $\{ |00\rangle ,|01\rangle,|10\rangle,|11\rangle \}$. 
In our case, it will prove convenient to employ the so called Bell basis of maximally correlated states, which are of the form

\begin{equation} \label{bellbasis}
|\Phi^{\pm} \rangle= \frac{(|00\rangle \pm e^{i\theta}|11\rangle)}{\sqrt{2}}, |\Psi^{\pm} \rangle= \frac{(|01\rangle \pm e^{i\theta}|10\rangle)}{\sqrt{2}}.
\end{equation}

\noindent Now we rise the question of wether all elements of $\rho$ intervene in the computation of  $B_{CHSH}$. Given a state $\rho$, we make a change of basis 
so that we work in the Bell basis. For simplicity, and without loss of generality, we shall take real coefficients ($\theta=0$ in (\ref{bellbasis})). 
Due to the special form for (\ref{CHSH_QM}), we separate the elements of $\rho$ (now in the Bell basis) into two contributions, namely

\begin{eqnarray} \label{decomp}
\rho = \rho_{\parallel} + \rho_{\perp}&=& \left( \begin{array}{cccc}
\rho_{11} & i\rho^{I}_{12} & i\rho^{I}_{13} & \rho^{R}_{14}\\
-i\rho^{I}_{12} & \rho_{22} & \rho^{R}_{23} & i\rho^{I}_{24}\\
-i\rho^{I}_{13} & \rho^{R}_{23} & \rho_{33} & i\rho^{I}_{34}\\
\rho^{R}_{14} & -i\rho^{I}_{24} & -i\rho^{I}_{34} & \rho_{44} \end{array} \right)  \\
&+&
\left( \begin{array}{cccc}
0 & \rho^{R}_{12} & \rho^{R}_{13} & i\rho^{I}_{14}\\
\rho^{R}_{12} & 0 & i\rho^{I}_{23} & \rho^{R}_{24}\\
\rho^{R}_{13} & -i\rho^{I}_{23} & 0 & \rho^{R}_{34}\\
-i\rho^{I}_{14} & \rho^{R}_{24} & \rho^{R}_{34} & 0 \end{array} \right).
\end{eqnarray}

\noindent This separation is motivated by the fact that only
terms in $\rho_{\parallel}$ contribute to $Tr(\rho B_{CHSH})$, as one can easily check. 
In other words, $Tr(\rho B_{CHSH})=Tr(\rho_{\parallel}  B_{CHSH})+Tr(\rho_{\perp}  B_{CHSH})=Tr(\rho_{\parallel}  B_{CHSH})$. 
The superscripts of the matrix elements in (\ref{decomp}) refer to the concomitant real ($R$) and imaginary ($I$) parts. This 
observation constitutes the starting point of our study.

The answer to the initial question of this Section involves only the elements of $\rho_{\parallel}$ for $\rho$  in (\ref{decomp}). The latter fact enormously simplifies 
the computation, but we nevertheless encounter a highly nontrivial optimization enterprise. 

Fortunately, we do not require all elements of  $\rho_{\parallel}$. Instead, since we seek maximum nonlocality, we 
will consider states which are diagonal in the Bell basis (null elements off-diagonal in $\rho_{\parallel}$ (\ref{decomp})), 
for nonlocal correlations concentrate after some depolarizing process \cite{depol}. 
Previous authors computed the entanglement and the maximum violation for the Bell inequality for 
Bell diagonal states \cite{basis1}, and the particular form for these states \cite{basis2}. In the present work, we re obtain and extend 
their results by means of a specific optimization technique, which is described in full detail in Appendix I.

For diagonal states in the Bell basis

\begin{eqnarray} \label{rhodiag}
\rho_{Bell}^{(diag)}&=& \lambda_1| \Phi^{+} \rangle \langle \Phi^{+}| +\lambda_2| \Phi^{-} \rangle \langle \Phi^{-}| \cr
&+&\lambda_3| \Psi^{+} \rangle \langle \Psi^{+}| +\lambda_4| \Psi^{-} \rangle \langle \Psi^{-}|,
\end{eqnarray}

\noindent with eigenvalues appearing in decreasing order, we obtain

\begin{equation} \label{resultat}
 \max_{\bf{a_j},\bf{b_j}} Tr (\rho_{Bell}^{(diag)} {\cal B}_{CHSH}) = 2\sqrt{2} \sqrt{(\lambda_1 - \lambda_4)^2 + (\lambda_2 - \lambda_3)^2}.
 \end{equation}

\noindent Recall that $2\sqrt{2}$ is the maximum value allowed by quantum mechanics (attained only for states (\ref{bellbasis})).

We are going to determine which is the maximum expectation value of the CHSH operator (\ref{CHSH_QM}) that a two qubit mixed state 
$\rho$ with some degree of mixedness, in this case given by the so called participation ratio $R=1/Tr(\rho^2)$, may have. 
Notice that no assumption is needed regarding the state being 
diagonal or not in the Bell basis. In order to solve the concomitant variational problem 
(and bearing in mind that $B_{CHSH}=Tr(\rho{\cal B}_{CHSH})$), let us first find the state that extremizes 
Tr($\rho^2$) under the constraints associated with a given value of $B_{CHSH}$, and the normalization of $\rho$.
This variational problem can be cast as

\begin{equation} \label{var}
\delta \big[  Tr(\rho^2) + \beta Tr(\rho{\cal B}_{CHSH}) - \alpha Tr(\rho)   \big]=0,
\end{equation}

\noindent where $\alpha$ and $\beta$ are appropriate Lagrange multipliers.

The solution of the above variational equation is given by

\begin{equation} \label{rhovar}
\rho=\frac{1}{2} \big[  \alpha I - \beta {\cal B}_{CHSH} \big],
\end{equation} 

\noindent with $I$ being the $4\times4$ identity matrix. The value of the Lagrange multiplier 
$\alpha$ is immediately obtained by the normalization requirement, with $\alpha=\frac{1}{2}$. 

From (\ref{rhovar}) we find ${\cal B}_{CHSH}$, multiply it by $\rho$ and apply the corresponding definition 
of $B_{CHSH}^{\max}$, taking into account that $Tr(\rho^2)=1/R$. We arrive at the result 

\begin{equation} \label{varR}
B_{CHSH}=Tr(\rho{\cal B}_{CHSH}) = -\frac{2}{\beta} \cdot \frac{4-R}{4R}.
\end{equation} 

By either squaring (\ref{rhovar}) and taking the trace according to the definition of $R$, or rather multiply it by ${\cal B}_{CHSH}$ 
(${\cal B}_{CHSH}$ is traceless) in order to get $B_{CHSH}$, both ways  lead to $\beta=-\frac{B_{CHSH}}{8}$. 
Combining either the former or the latter result with relation (\ref{varR}), we finally \cite{explicacio} arrive at

\begin{equation} \label{varResult}
B_{CHSH}^{\max}=\sqrt{Tr[{\cal B}_{CHSH}^2]} \cdot \sqrt{ \frac{4-R}{4R} } = 4\cdot \sqrt{ \frac{4-R}{4R} }. 
\end{equation} 

\noindent This result is valid for the range $R\in[2,4]$. The corresponding state (\ref{rhovar}) can now be cast in the 
new form

\begin{equation} \label{state2}
\rho_{II}=diag\bigg(x,x,\frac{1-2x}{2},\frac{1-2x}{2}\bigg),
\end{equation} 

\noindent with $x\in [0,\frac{1}{4}]$ and diagonal in the Bell basis. In the region $R\in[1,2]$ the form of state (\ref{state2}) is no longer 
valid. Instead, we look for those states that stay close to pure states (maximum nonlocality) and possess rank 2 (following 
the requirement $R\in[1,2]$). The simplest case is that of a state diagonal in the Bell basis, being of the form 

\begin{equation} \label{state1}
\rho_I=diag(1-x,x,0,0)
\end{equation} 

\noindent with $x\in [0,\frac{1}{2}]$. Relation (\ref{resultat}) returns 

\begin{equation} \label{region1}
B_{CHSH}^{\max}(\rho_I)= 2\sqrt{2} \sqrt{(1-x)^2 + x^2} = \sqrt{\frac{8}{R}}.
\end{equation} 

We are now in a position to answer the initial question of this Section. Besides, we do not find the functional 
form for $B_{CHSH}^{\max}(R)$ for two qubits, we do also obtain the form for those states, states which are diagonal in the Bell basis. 
We shall call these states Maximally Nonlocal Mixed States (MNMS). 

For a given value of the participation ratio $R$, we can obtain a more general class of states rather than $\rho_I$ (\ref{state1}) and $\rho_{II}$ (\ref{state2}) 
by letting a non-zero phase $\theta$ in (\ref{bellbasis}). By doing so, and rewriting the concomitant MNMS states $\rho_I$ (\ref{state1}) in the computational basis, 
one easily obtains the family of states provided in \cite{basis2}. But not only this: a whole series of new states possessing maximum nonlocality for a 
given value of $R$ are obtained by changing the position of the eigenstates of $\rho_I$ in (\ref{state1}). This is possible since no preferred disposition of states in the Bell basis 
is required for diagonal states (\ref{rhodiag}) as far as maximum amount of nonlocality is concerned.  

Let us summarize all previous results: {\it the maximum amount of nonlocality attained by a mixed state $\rho$ of two qubits is 
given by}

\begin{itemize}
\item Mixedness described by the participation ratio $R=1/Tr(\rho^2)$

\begin{equation}\label{resumR}
\begin{split}
B_{CHSH}^{\max}(R)&=\sqrt{\frac{8}{R}} , R\in[1,2]\\ 
B_{CHSH}^{\max}(R)&= 4\cdot \sqrt{ \frac{4-R}{4R} }, R\in[2,4]
\end{split}
\end{equation}

\item Mixedness described by the maximum eigenvalue $\lambda_{\max}(\rho)$

\begin{equation}\label{resumL}
\begin{split}
\frac{B_{CHSH}^{\max}(\lambda_{\max})}{2 \sqrt{2}}&=  (4\lambda_{\max}-1), \lambda_{\max}\in \bigg[\frac{1}{4},\frac{1}{3}\bigg]\\ 
\frac{B_{CHSH}^{\max}(\lambda_{\max})}{2 \sqrt{2}}&= \sqrt{\lambda_{\max}^2+(1-3\lambda_{\max})^2}, \lambda_{\max}\in\bigg[\frac{1}{3},\frac{1}{2}\bigg]\\ 
\frac{B_{CHSH}^{\max}(\lambda_{\max})}{2 \sqrt{2}}&= \sqrt{\lambda_{\max}^2+(1-\lambda_{\max})^2}, \lambda_{\max}\in\bigg[\frac{1}{2},1\bigg] 
\end{split}
\end{equation}

\end{itemize}

One must bear in mind that states that reach the maximum possible value for nonlocality measure $B_{CHSH}^{\max}$ greatly depend 
on what measure for the degree of mixture is employed. The only case where both descriptions agree is for those states that strictly violate 
the CHSH inequality, namely, the MNMS $\rho_I$ (\ref{state1}).

\subsection{Nonlocality for maximally entangled mixed states}


Maximally entangled mixed states (MEMS) constitute a family of states that are 
maximally entangled for a given degree of mixture, 
measured by the participation ratio $R=1/Tr(\rho^2)$. In practice, one will
more often have to deal with mixed states than with pure ones.
From the point of view of entanglement-exploitation, one should
then be interested in MEMS states $\rho_{MEMS}$, which are basic 
constituents of quantum communication protocols. The MEMS states have been studied, for example,
in Refs. \cite{MJWK0,W03,BatleEntropy}. MEMS states have been experimentally 
encountered \cite{memsexp,memsexp2}.
In the computational basis $\{ |00\rangle ,|01\rangle,|10\rangle,|11\rangle \}$, 
they are written as

\begin{equation} \label{MEMScompu}
\left( \begin{array}{cccc}
g(x) & 0 & 0 & x/2\\
0 & 1 - 2g(x) & 0 & 0\\
0 & 0 & 0 & 0\\
x/2 & 0 & 0 & g(x)
\end{array} \right),
\end{equation} 

\noindent with $g(x)=1/3$ for $0\le x \le 2/3$, and  $g(x)=x/2$ for $2/3 \le x
\le 1$. The quantity $x$ is equal to the concurrence $C$. 
The change of $g(x)-$regime ensues for $R=1.8$.

Our goal is to uncover interesting correlations between
entanglement, nonlocality and mixedness that emerge for these states. 
Indeed, the study of the nonlocality of these states offers an excellent framework 
where to compare the extremal cases for nonlocality and entanglement. 
The MEMS states (\ref{MEMScompu}) are written in the Bell basis 
$\{ |\Phi^{+}\rangle,|\Phi^{-}\rangle,|\Psi^{+}\rangle,|\Psi^{-}\rangle \}$ in the form

\begin{equation} \label{MEMSBel}
\left( \begin{array}{cccc}
g(x)+\frac{x}{2} & 0 & 0 & 0\\
0 & g(x)-\frac{x}{2} & 0 & 0\\
0 & 0 & \frac{1-2g(x)}{2} & \frac{1-2g(x)}{2}\\
0 & 0 & \frac{1-2g(x)}{2} & \frac{1-2g(x)}{2} 
\end{array} \right),
\end{equation} 

\noindent which is to be compared with the general form for arbitrary states (\ref{decomp}) in the Bell basis. The direct comparison 
of MEMS states in the Bell basis yields to the conclusion that states 
$\rho_{MEMS}$ (\ref{MEMSBel}) {\it behave as if they were diagonal in the Bell basis as far as nonlocality is concerned}.

This crucial observation allow us to the simple calculation of the maximum amount of nonlocality for MEMS states to be 
of the form

\begin{equation} \label{BellMEMS}
B_{CHSH}^{\max} (x) = \begin{cases} \frac{2}{3}\sqrt{1+9x^2}, \,\,\,\, 0\leq x \leq \frac{1}{3}\\ 
2\sqrt{2}x, \,\,\,\,\,\,\,\,\,\,\,\,\, \,\,\,\, \frac{1}{3} < x \leq 1 \end{cases} 
\end{equation} 

\noindent Recall that $x=C$, the so called concurrence. Also, it is plain from relation (\ref{BellMEMS}) that any bipartite state 
possessing $x\geq \frac{1}{\sqrt{2}}$ will violate the CHSH inequality, since no state is more entangled that the MEMS states. 

The concurrence entanglement indicator $C$ for MNMS states $\rho_I$ (\ref{state1}) can be easily calculated to be $C=1-2x$. Therefore, 
the relation between nonlocality $B_{CHSH}^{\max}$ and $C$ is such that $B_{CHSH}^{\max}=2\sqrt{1+C^2}$, which easily recovers 
a previous result  \cite{basis1}. In other words, for a given value of $C$, MNMS states possess the maximum possible violation of the CHSH inequality 
while, on the contrary, MEMS possess a minimum amount of CHSH violation ($2\sqrt{1+C^2} > 2\sqrt{2}C$). 

As a consequence, we draw the conclusion that 
maximum entanglement for mixed states of two qubits does not imply maximum nonlocality, though in this extremal case nonlocality and entanglement are 
monotonic increasing functions one of the other.

\subsection{A physically-motivated case: the $XY$ model}

The general study of nonlocality and entanglement in an infinite quantum system was performed in Ref. \cite{noltros}. 
In this section we incorporate new results and generelize previous ones in the light of 
the bounds encountered for the maximum violation of the CHSH Bell inequality for a given degree of mixture. 

The general two-site density matrix for two spins along the $XY$ chain is expressed as

\begin{equation} \label{rho2}
\rho_{ij}^{(R)} = \frac{1}{4} \,
\Bigg[ \mathbb{I} + \sum_{u,v} T_{uv}^{(R)}
\sigma^i_u \otimes \sigma^j_v  \Bigg].
\end{equation}

\noindent $R=j-i$ indicates the distance between spins (not to be confused with the participation ratio $R=1/Tr(\rho^2)$, $\{u,v\}$ denote any index of 
$\{\sigma_0,\sigma_x,\sigma_y,\sigma_z\}$, and $T_{uv}^{(R)} \equiv \langle \sigma^i_u \otimes \sigma^j_v \rangle$. 
Due to symmetry considerations, only
$\{T_{xx}^{(R)},T_{yy}^{(R)},T_{zz}^{(R)},T_{xy}^{(R)}\}$ do not vanish. Barouch {\it et al} \cite{Barouch} provided
exact expressions for two-point correlations, together with all the dynamics associated with an 
external magnetic field $h(t)$ along the $z$-axis. We shall consider the case where $h$ jumps from and initial value $h_0$ to a final
value $h_f$ at $t=0$, that is, a quench (the equilibrium case is easily recovered when $h_f=h_0$) 
and the $R=1$ configuration (nearest neighbors). 

The most remarkable result of Ref. \cite{noltros} as far as nonlocality is concerned is that the maximum value for the 
quantity $B_{CHSH}^{\max}$ for states (\ref{rho2}), given by twice the expression

\begin{equation} \label{CHSHmax}
 \sqrt{ \Vert {\bf T^{(R)}} \Vert^2 - \min \big( \big[ T_{xx}^{(R)} \big]^2,
\big[ T_{yy}^{(R)} \big]^2, \big[ T_{zz}^{(R)} \big]^2 \big) + 2 \big[ T_{xy}^{(R)} \big]^2},
\end{equation}

\noindent with ${\bf T^{(R)}}=(T_{xx}^{(R)},T_{yy}^{(R)},T_{zz}^{(R)})$, is always $\leq 2$ 
{\it for any configuration $R$ and any non-zero value of the entanglement},

Fig. 1 depicts several time evolution for the state (\ref{rho2}) once a quench in the external magnetic field is applied. 
As a consequence, we have a nonergodic evolution in time, which translates into oscillating values for both 
$B_{CHSH}^{\max}$ and $R$. The previous time dependent cases correspond to Fig. 1(a) and Fig. 1(b). 
The static case is depicted in Fig. 1(c), where nonlocality improves for states approaching the 
Ising case ($\gamma=1$). Several time evolution plots appear in Fig. 1(d) for $B_{CHSH}^{\max}$ and the entanglement 
of formation for states (\ref{rho2}), together with the thermodynamic magnetization $M_z$ after a quench 
from $h_0=0.5$ to $h_f=0$. All these three quantities possess a nonergodic behavior, which is not surprising for 
they ultimately depend on two point spin correlators, which in turn are nonergodic quantities \cite{noltros}.

\begin{figure}[htbp]
\begin{center}
\includegraphics[width=8.6cm]{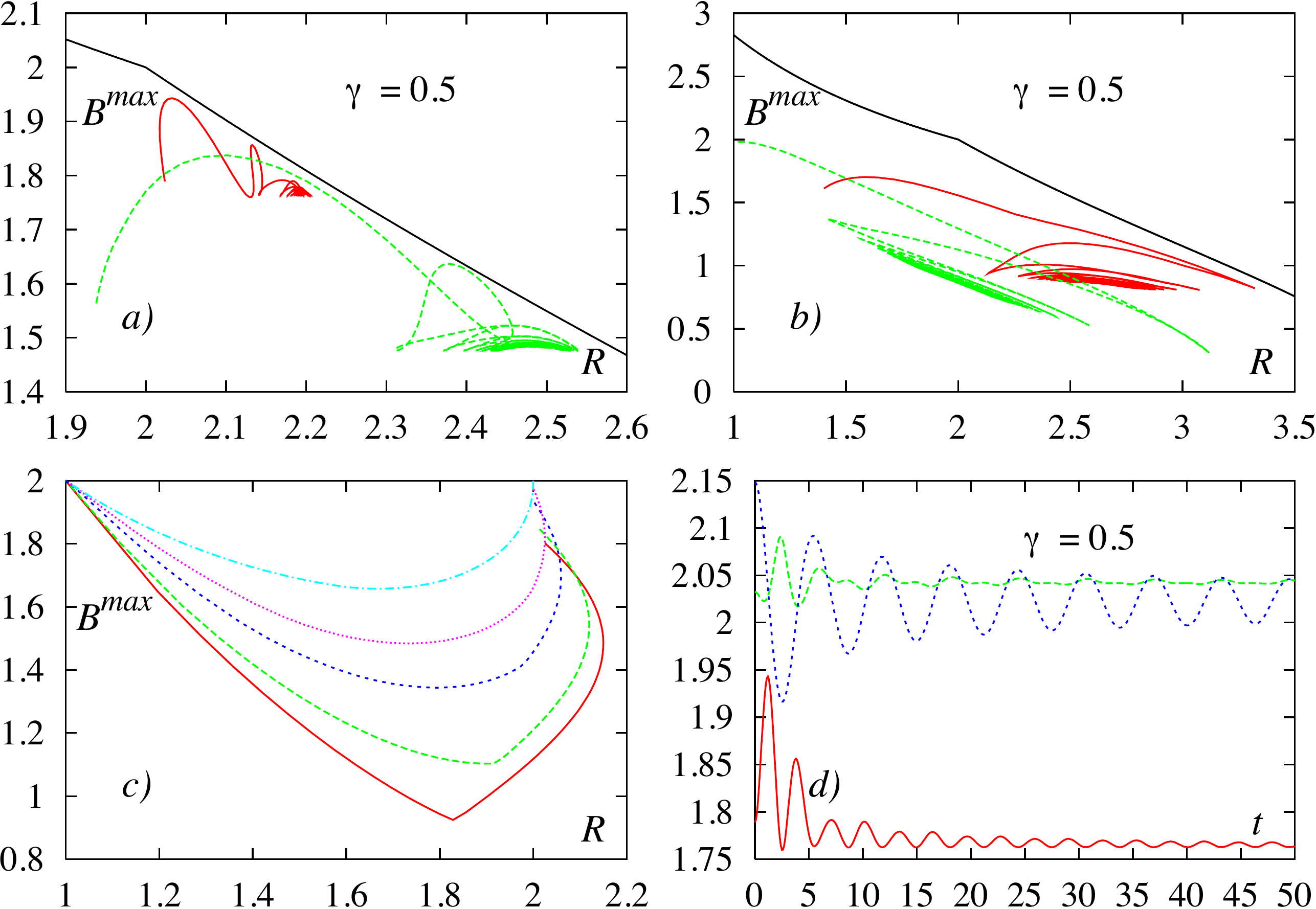}
\caption{(Color online) (a) Value of $B_{CHSH}^{\max}$ vs $R$ for several time 
evolutions of the external perpendicular magnetic field in the $XY$ model. The solid curve corresponds to the case 
$(h_0=0.5,h_f=0)$. From left to right, both $B_{CHSH}^{\max}$ and $R$ oscillate around their concomitant final 
(non-equilibrium) values. The lower dashed curve depicts the case $(h_0=0.75,h_f=0)$, with a similar behavior. 
Notice that in either cases the maximum value of $B_{CHSH}^{\max}(R)$ (upper solid line) is never crossed. 
(b) Similar curves for $(h_0=1,h_f=0)$ (solid curve) and $(h_0=2,h_f=0)$ (lower dashed curve). 
(c) $B_{CHSH}^{\max}$ vs $R$ plots for several anisotropy values (from bottom to top) 
$\gamma=0,0.1,0.3,0.5,1$. As the magnetic field 
$h$ increases from 0 to $\infty$ (no time evolution), the curves go from right to left 
(decreasing $R$-values for the two qubit states). It is plain from this series of plots that 
no violation of the CHSH Bell inequality occurs. 
(d) Time evolution of $B_{CHSH}^{\max}$ (lower solid curve), entanglement $E$ (long dashed curve) and the magnetization $M_z$ 
(short dashed curve) after the quench $(h_0=0.5,h_f=0)$. $E$ and $M_z$ have been shifted two units upwards. 
All these three quantities are nonergodic. See text for details.}
\label{fig1}
\end{center}
\end{figure}

In Fig. 2, we consider the maximum value for $B_{CHSH}^{\max}$ and the concurrence $C$ that any state of the type (\ref{rho2}) can have. 
The value for the factorizing field $h_s=\sqrt{1-\gamma^2}$ for which states (\ref{rho2}) are separable \cite{Barouch} correspond to zero 
concurrence, that is, a line at the bottom of the plot. In the language of mixedness, the magnetic field $h$ and the mixture of the 
state go in opposite directions: the greater the former, the lesser the latter. This is so because as $h \rightarrow \infty$, we approach 
a pure state ($R=1$) with both spins down. In the surface $\gamma=$ constant we encounter that both $B_{CHSH}^{\max}$ and $C$ diverge (their 
first derivative with respect to $h$), thus signaling a quantum phase transition (except for 
the isotropic case $\gamma=0$).

\begin{figure}[htbp]
\begin{center}
\includegraphics[width=8.6cm]{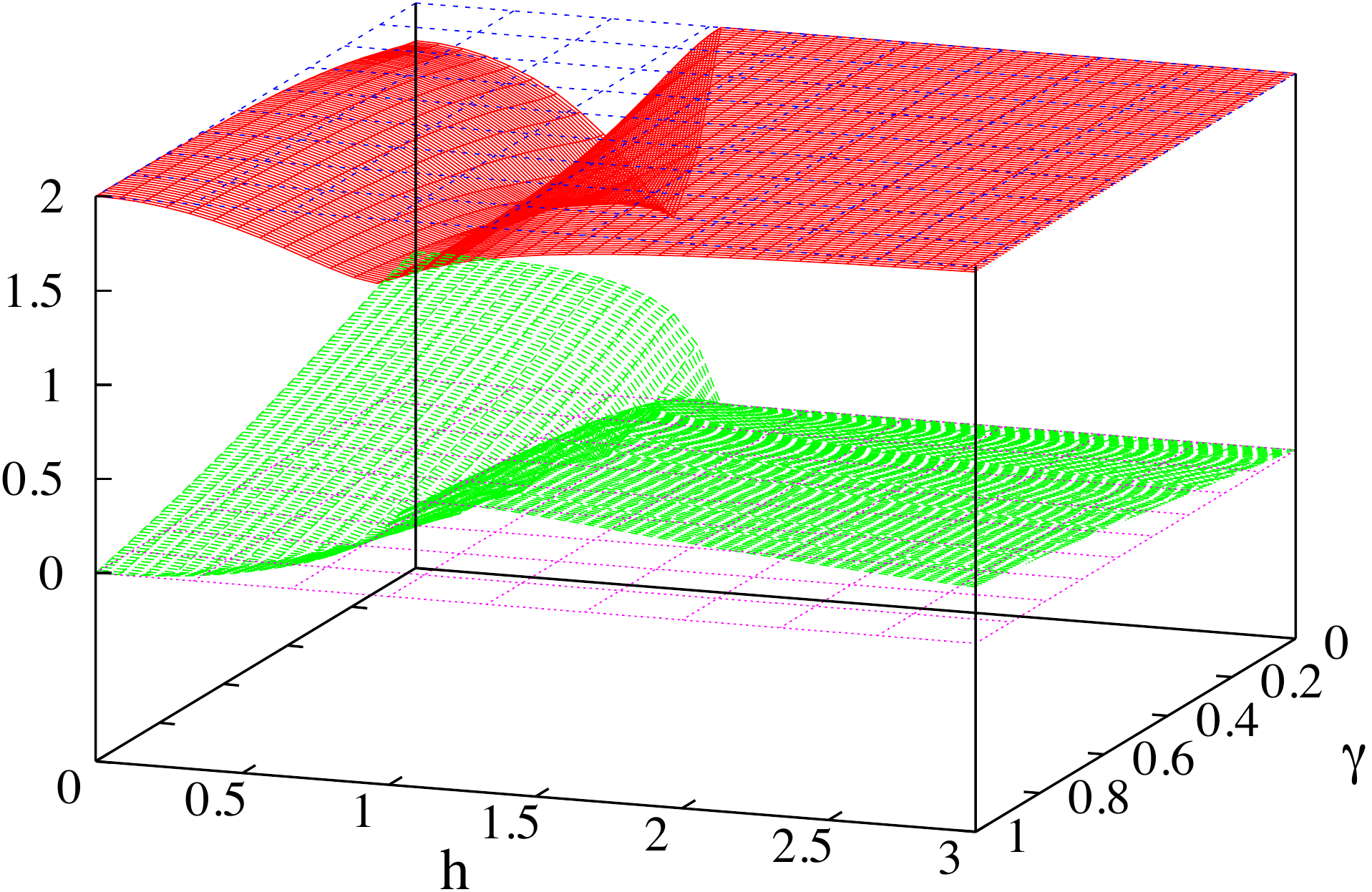}
\caption{(Color online) Plot of the nonlocality measure $B_{CHSH}^{\max}$ (upper surface) and 
twice the concurrence $C$ (lower surface) for any value of the anisotropy $\gamma$ and the 
external perpendicular magnetic field $h$ for two qubit states (nearest neighbors) in the XY model. 
It is plain that no violation of the CHSH Bell inequality occurs, regardless of the non-zero value for 
the entanglement indicator $C$. See text for details.}
\label{fig3}
\end{center}
\end{figure}

As we can see, the notion of nonlocality and entanglement as different resources appear in a real physical 
system. The necessity of the violation of some Bell inequality for some information-theoretic tasks and its relation to entanglement 
makes the whole picture a bit more intriguing with this physical case.

\subsection{Nonlocality vs entanglement for linear combinations of pure states of two qubits}


Nonlocality may also exhibit interesting features for pure states. In our case we shall consider states of the form 

\begin{equation} \label{Bellpure}
|\phi\rangle = \sqrt{\lambda_1}|\Phi^{+}\rangle + \sqrt{\lambda_2}|\Phi^{-}\rangle 
+ \sqrt{\lambda_3}|\Psi^{+}\rangle + \sqrt{\lambda_4} |\Psi^{-}\rangle,    
\end{equation} 

\noindent with real coefficients $\{ \sqrt{\lambda_i} \}$. This particular form clearly constitutes an extension of the analysis performed for 
two qubit mixed states diagonal in the Bell basis. All the details of the optimization of the CHSH inequality for states (\ref{Bellpure}) is 
given in Appendix I. The final result is 

\begin{equation}\label{Bellresult}
2\sqrt{2} \sqrt{(\lambda_1 + \lambda_4)^2 + (\lambda_2 + \lambda_3)^2}.
\end{equation}

We know by virtue of Gisin's theorem \cite{Gisin} that pure bipartite entanglement implies violation of the CHSH inequality. The question is wether 
that dependency changes when we consider the case of linear combinations. The corresponding answer is no, because otherwise 
it would imply a preferred choice for the state basis. 

The value (\ref{Bellresult}) for maximum violation of the CHSH inequality is to be compared with the measure of entanglement 
for the state (\ref{Bellpure})

\begin{equation} \label{BellConc}
C^2=4\,\det \,\rho_{A/B}=1-4(\lambda_1+\lambda_4)(\lambda_2+\lambda_3).
\end{equation} 

\noindent Strictly speaking, the concurrence or its squared value C$^2$ are not proper measures of entanglement, 
for they do not comply with the usual basic requirements \cite{LoPo}. 
However, they are widely used as useful entanglement quantifies (they are monotonic functions of the entanglement of 
formation $E_f(\cdot)$ \cite{Wootters}, which is a good measure).  

Taking into account the form for $C^2$ (\ref{BellConc}), and the value for ${\cal B}_{CHSH}^{\max}$ (\ref{Bellresult}), we derive the 
relation

\begin{equation} \label{Bellrelation}
{\cal B}_{CHSH}^{\max} = \sqrt{4 C^2 + 4},
\end{equation}

\noindent which is the same result we would obtain for a pure state of two qubits written in the Schmidt basis. Had we assumed 
result (\ref{Bellrelation}) to hold for any state, we would have obtained the relation (\ref{Bellresult}) for the maximum violation of the CHSH 
inequality without recourse to any optimization technique.  

Formula (\ref{Bellresult}) has interesting echoes when compared to the entanglement of the same superposition of states. 
Entanglement of superposition of states was originally conceived in Ref. \cite{superpos}, where interesting bounds for the superposed 
state were obtained in terms of their constituents. In our case, result (\ref{Bellresult}) permits us to establish similar bounds for 
maximum violation of the CHSH inequality. This framework offers a link between the characterization of nonlocality and entanglement, 
where their mutual intricacies become more apparent. 

As obtained before, we know by virtue of Gisin's theorem \cite{Gisin} that nonlocality implies entanglement (and vice versa) for pure two qubit states, 
but nothing is said regarding their particular characterization. Let us illustrate, before embarking on our study, what happens when we consider the nonlocality 
present in the superposed state

\begin{equation} \label{Bellstate}
|\theta\rangle = \alpha |01\rangle + \beta |\Phi^{+}\rangle,
\end{equation}

\noindent with ${\cal B}_{CHSH}^{\max}(|01\rangle)=2$ and ${\cal B}_{CHSH}^{\max}(|\Phi^{+}\rangle)=2\sqrt{2}$. Presumedly, 
nonlocality of state $|\theta\rangle$ should be lowered by the action of non correlated $|01\rangle$. Indeed, we have

\begin{equation} \label{Bellexpand}
{\cal B}_{CHSH}^{\max}(|\theta\rangle)=2\sqrt{2-\alpha^2(2-\alpha^2)}=2\sqrt{2}-\sqrt{2}\alpha^2 + O(\alpha^4).
\end{equation}

\noindent This example shows that ${\cal B}_{CHSH}^{\max}$ and the fidelity between states within the linear combination simultaneously and continuously change, 
a fact that does not occur for the entanglement of linear combination of states  \cite{superpos}.

{\it Theorem 1:} Given a set of $M$ orthogonal pure states of two qubits $\{ |\phi_i \rangle \}_{i=1}^M$, with $\sum_{i=1}^M \alpha_i^2 =1 \,(\alpha_i \in \mathbb{R})$, 
the concomitant maximum nonlocality measure obeys

\begin{equation} \label{Bellgeneral}
{{\cal B}_{CHSH}^{\max}}^2 \bigg( \sum_{i=1}^M \alpha_i  |\phi_i \rangle \bigg)  \geq   \sum_{i=1}^M (\alpha_i)^4 {{\cal B}_{CHSH}^{\max} }^2 (|\phi_i\rangle ).
\end{equation}

{\it Proof:} Spanning the set of states $\{ |\phi_i \rangle \}_{i=1}^M$ in the Bell basis, by recourse to (\ref{Bellresult}) and 
expanding quadratic terms, the remaining part on the right hand side of is a strictly positive quantity, from hence we directly compute
the nonlocality of superposition as stated in Theorem 1. The upper bound for ${{\cal B}_{CHSH}^{\max}}^2 \bigg( \sum_{i=1}^M \alpha_i  |\phi_i \rangle \bigg)$ 
is easily obtained by individually optimizing each term in the argument $\{ |\phi_i \rangle \}$, and taking into account each contribution arising from 
2$Re\big[  {\cal B}_{CHSH}^{\max} (\langle \phi_i  |\phi_j \rangle ) \big],\, \forall \,i\neq j$.

Theorem 1 and the concomitant upper bound connect the way nonlocality of a superposed state is distributed among its constituents, in a similar  fashion as entanglement 
in Ref. \cite{superpos}. More details on superposition of states will be described elsewhere \cite{futurLin}.

\section{THREE QUBITS}

\subsection{Nonlocality for three qubit states. Application to the $XY$ model}

We shall explore nonlocality in the three qubit case through the violation of the Mermin inequality \cite{Mermin}. 
This inequality was conceived originally in order to detect genuine three-party quantum correlations impossible to reproduce 
via LVMs. The Mermin inequality reads as $Tr(\rho {\cal B}_{Mermin}) \leq 2$, where ${\cal B}_{Mermin}$ is the Mermin operator

\begin{equation} \label{Mermin}
 {\cal B}_{Mermin}=B_{a_{1}a_{2}a_{3}} - B_{a_{1}b_{2}b_{3}} - B_{b_{1}a_{2}b_{3}} - B_{b_{1}b_{2}a_{3}},
\end{equation}

\noindent with $B_{uvw} \equiv {\bf u} \cdot {\bf \sigma} \otimes {\bf v} \cdot {\bf \sigma} \otimes {\bf w} \cdot {\bf \sigma}$ 
with ${\bf \sigma}=(\sigma_x,\sigma_y,\sigma_z)$ being the usual Pauli matrices, and ${\bf a_j}$ and ${\bf b_j}$ unit vectors
in $\mathbb{R}^3$. Notice that the Mermin inequality is maximally violated by Greenberger-Horne-Zeilinger (GHZ) states.  
As in the bipartite case, we shall define the following quantity

\begin{equation} \label{MerminMax}
 Mermin^{\max} \equiv \max_{\bf{a_j},\bf{b_j}}\,\,Tr (\rho {\cal B}_{Mermin})
\end{equation}

\noindent as a measure for the nonlocality of the state $\rho$. While in the bipartite the CHSH inequality 
was the strongest possible one, this is not the case for three qubits. The Mermin inequality is not the only existing Bell inequality for 
three qubits, but it constitutes a simple generalization of the CHSH one to the tripartite case. Therefore, it will suffice to use 
this particular inequality to illustrate the basic results of the present work.  

In view of the previous definitions, we are naturally led to the question of what class of three qubit mixed states possesses a maximum 
amount of nonlocality (\ref{MerminMax}), how does it look like and what it amounts for. Let us recall that the family of pure states of three qubits
$|\Psi_j^{ \pm } \rangle =(|j\rangle \pm |7-j\rangle)/\sqrt{2}$ forms a basis, the so called GHZ basis or Mermin basis, and that 
these states maximally violate the Mermin inequality. But what is the state of affairs for the general, mixed case? Given a state 
$\rho$, it can always be transformed into the state

\begin{equation} \label{rhodiag}
\rho_{Mermin}^{(diag)}= \sum_{j=0}^3 (\lambda_j^{+}|\Psi_j^{+} \rangle \langle \Psi_j^{+}| + \lambda_j^{-}|\Psi_j^{-} \rangle \langle \Psi_j^{+}|),
\end{equation}

\noindent a state which is diagonal in the Mermin basis, for nonlocal correlations concentrate after the action of some depolarizing process \cite{depol}. 
Without loss of generality, we can assume the eigenvalues of (\ref{rhodiag}) to be sorted in decreasing order, 
that is, $\lambda_0^{+} \geq \lambda_0^{-} \geq \dots \geq \lambda_3^{+}  \geq \lambda_3^{-}$, 
since otherwise it could be adjusted by a local unitary operation. 

The details of the optimization are given in Appendix II. However, the maximum violation of the Mermin inequality is given by the quantity

\begin{equation}\label{MerminMax2}
Mermin^{\max} \leqq 4\sqrt{\sum_{j=0}^3 (\lambda_j^{+} -\lambda_j^{-})^2}.
\end{equation}

\noindent The exact form for $Mermin^{\max}$ is rather unpleasant. In practice, the previous bound is an excellent one, differing from the exact one by a small 
amount and being equal in those cases where we have a high degree of symmetry in the state. For most practical purposes, one can consider the equality 
in (\ref{MerminMax2}) to hold.
\newline

We can encounter too interesting nonlocality features if we focus our attention to the case of pure states of three qubits being linear combinations in the Mermin 
basis $|\Psi_j^{ \pm } \rangle =(|j\rangle \pm |7-j\rangle)/\sqrt{2}$. That is, states of the form  

\begin{equation} \label{lincombpure}
|\phi\rangle = \sum_{j=1}^{8} \sqrt{\lambda_j^{\pm}} |\Psi_j^{\pm}\rangle,    
\end{equation} 

\noindent with real coefficients $\{ \sqrt{\lambda_j^{\pm}} \}$ such that $\sum_{j=1}^{8} \lambda_j^{\pm}=1$. 
\newline

Detection and characterization of entanglement in multipartite systems constitutes a hot research topic
in QIT. However, no necessary and sufficient criterion is available to date that discriminates
wether a given state of a multipartite system is entangled or not. 
Indeed, highly entangled multipartite states raise enormous interests in quantum information
processing and one-way universal quantum computing \cite{Briegel}. They are essential for several
quantum error codes and communication protocols \cite{Cleve}, as they are robust against decoherence.

In spite of the previous unbalanced present status between entanglement and nonlocality, the relation between both quantities for three qubits is seen in a 
new light when we study both resources for those states that attain the maximum possible nonlocality value given by expression (\ref{MerminMax2}). 
One way is to consider what class of particular states (\ref{rhodiag}) is maximally nonlocal for 
a given value of their degree of mixture, which is a tool employed to characterize mixed states. 
If we choose the participation ratio $R=1/Tr(\rho^2)$, we can obtain what is the functional form of  (\ref{MerminMax2}) in terms of $R$.

This procedure is virtually identical to the variational calculation performed for two qubit mixed states. Following the exact treatment as in Eq. (\ref{var}), 
and taking into account that $Tr[{\cal B}_{Mermin}^2]=32$, we obtain  

\begin{equation}\label{MerminR}
Mermin^{\max} (R) \propto \sqrt{\frac{8-R}{8R}}.
\end{equation}

\noindent The constant in (\ref{MerminR}) is obtained by requiring $Mermin^{\max}$ to be equal to 4 for pure states ($R=1$). 
One class of states that possess 
the previous optimal value is $\rho^{diag}=(1-7x,x,x,x,x,x,x,x)$, which is, interestingly enough, the generalized Werner state for three qubits 

\begin{equation}\label{MerminW}
\rho_W^{n=3}= \tilde{x}|GHZ\rangle \langle GHZ| + \frac{1-\tilde{x}}{8} I_8, 
\end{equation}

\noindent where $I_8$ is the $8\times8$ identity matrix, and $\tilde{x}=1-8x$. Notice that this was not the case for the two qubit instance. 

This interesting feature enables us to discuss the different ranges for $R$ where 
to compare nonlocality and presence of genuine tripartite entanglement. On the one hand, from (\ref{MerminR}) we obtain the nonlocality 
critical value $R_1=32/11\approxeq 2.9$: {\it no three qubit states possess any nonlocality for participation ratios $R \geq R_1$}. On the other hand, 
the special nature of generalized Werner states allow us to compute the separability threshold between entanglement 
and separability \cite{WernerThreshold}. From Ref. \cite{WernerThreshold}, the contribution $\tilde{x}$ in (\ref{MerminW}) is such that $x\leq1/5$ involves absence of 
entanglement. Translated into $R$-language, it implies a second critical value $R_2=25/4=6.25$: 
{\it no three qubit states possess entanglement for participation ratios $R \geq R_1$}. 

Therefore, the range of $R$-values splits into three regions: i) between 1 (pure states) and $R_1$, maximum amounts of nonlocality imply the presence of entanglement; 
ii) between $R_1$ and $R_2$ we have no violation of the Mermin inequality, yet there exists entanglement; finally, iii) region between $R_2$ and $R=8$ 
(maximally mixed state) displays absence of both magnitudes. Notice, however, that in appearance there is some room left for LVM to hold in the second region, 
where no violation of the Mermin inequality occurs.
\newline

The exploration of nonlocality and entanglement for three qubit states would be incomplete without an specific example of their applicability, such as 
the possible information-theoretic tasks limitations imposed by the former on the latter. There is one such case, which is the well-known infinite $XY$ model 
in a transverse magnetic field \cite{Barouch}. This instance was explored in detail in Ref. \cite{noltros}. 

The $XY$ model is completely solvable, a fact that allows us to compute the reduced density matrix
for three spins without the explicit construction of the global infinite state of the system. 
The reduced state of three spins reads as

\begin{equation} \label{rho3}
\rho_{ijk}^{(a,b)} = \frac{1}{8} \,
\Bigg[ \mathbb{I} + \sum_{u,v,w} T_{uvw}^{(a,b)}
\sigma^i_u \otimes \sigma^j_v \otimes \sigma^k_w \Bigg],
\end{equation}

\noindent where $i<j<k$ indicate the positions of the three spins and $a=j-i, b=k-j$ their relative distances.
$\{u,v,w\}$ denote indexes of the Pauli matrices $\{\sigma_0,\sigma_x,\sigma_y,\sigma_z\}$,
and $T_{uvw}^{(a,b)} \equiv \langle \sigma^i_u \otimes \sigma^j_v \otimes \sigma^k_w \rangle_{ab}$. 
The calculation of the three-spin correlations $T_{uvw}^{(a,b)}$ were computed in Ref. \cite{noltros} by using the 
Wick theorem in quantum field theory. 

The most significant result is that, for any value of the anisotropy and external magnetic field, we have $Mermin^{\max}$ 
to be less than or equal to

\begin{equation} \label{result}
 \sqrt{4\big(T_{zzz}^{(a,b)}\big)^2 + 4\big(T_{zxx}^{(a,b)}\big)^2 + 4\big(T_{xzx}^{(a,b)}\big)^2 + 4\big(T_{xxz}^{(a,b)}\big)^2 },
\end{equation}

\noindent which is always $\leq 2$ for {\it any configuration of the spins (a,b) yet there is no null entanglement}. This constitutes a clear sign that 
states (\ref{rho3}) never violate the Mermin inequality, which entails an inherent limitation to the usefulness of entanglement itself. Furthermore, 
these states are shown to be distillable in most of the cases, which is a novel result: we have three-party distillable states in the $XY$ model 
with no violation of the Mermin inequality. The distillability issue constitutes the subject of the next section.

\subsection{Distillability and nonlocality for three qubit states}

Distillability and the violation of Bell inequalities --nonlocality-- constitute two manifestations of entanglement. While the former is related to the usefulness 
in quantum information processing tasks, due to the fact that most of them require pure-state entanglement as a key ingredient, the latter 
expresses the fact that a state cannot be simulated by classical correlations. In this vein, Gisin relates both resources when he points out 
in Ref \cite{Gisinnonlo} the question of wether there exists any bound entangled state that violates some Bell inequality. By bound entangled 
state one implies a state that cannot be distilled by means of local operations and classical communications. 
In the bipartite case, it has been shown \cite{Masanes} that no bound state violates the CHSH inequality. 


The separability criteria borrowed from the bipartite case, which employ positive partial transposition \cite{PPT} for all
parties, are all approximate. Nevertheless, this criterion based on the positivity of the ensuing partially
transposed matrix $\rho^{T_j}$ has a very interesting application. Notice that
if a three qubit state $\rho$ has positive $\rho^{T_j}$ for all $j=1,2,3$, where $T_j$
represents partial transposition for the system $j$, then it is said that the
state is GHZ-distillable, that is, one can distill a GHZ state from many copies
of $\rho$ by LOCC \cite{DurCiracTarrach}.

One of of questions that we want to address is wether there exists any nonlocal bound entangled state of three qubits. 
After applying a series of local transformations, one can convert any state into one belonging to the family 
$\rho_{Mermin}^{(diag)}$ (\ref{rhodiag}). 
For a state of three qubits to be non-distillable (bound entangled), any of the subsequent following inequalities must hold:


\begin{equation}\label{ineq}
\begin{split}
\rho^{T_1} > 0 &\Rightarrow \begin{cases} \lambda_2^{+} +\lambda_2^{-} > \lambda_1^{+} -\lambda_1^{-}\\ 
\lambda_3^{+} +\lambda_3^{-} > \lambda_0^{+} -\lambda_0^{-} \end{cases} \Rightarrow \lambda_0^{+} +\lambda_1^{+} < \frac{1}{2},\\
\rho^{T_2} > 0 &\Rightarrow \begin{cases} \lambda_2^{+} +\lambda_2^{-} > \lambda_0^{+} -\lambda_0^{-}\\ 
\lambda_3^{+} +\lambda_3^{-} > \lambda_1^{+} -\lambda_1^{-} \end{cases} \Rightarrow \lambda_0^{+} +\lambda_1^{+} < \frac{1}{2},\\
\rho^{T_3} > 0 &\Rightarrow \begin{cases} \lambda_1^{+} +\lambda_1^{-} > \lambda_0^{+} -\lambda_0^{-}\\ 
\lambda_3^{+} +\lambda_3^{-} > \lambda_2^{+} -\lambda_2^{-} \end{cases} \Rightarrow \lambda_0^{+} +\lambda_2^{+} < \frac{1}{2},
\end{split}
\end{equation}

\noindent where the last inequality in each case is a consequence of the sum of the previous two. None of the previous inequalities for the 
eigenvalues of states diagonal in the Mermin-basis is compatible with (\ref{MerminMax}) being greater than 2, which implies that {\it no bound entangled state is 
present in Mermin-diagonal mixed states of three qubits that violates the Mermin inequality}. 

\begin{figure}[htbp]
\begin{center}
\includegraphics[width=8.6cm]{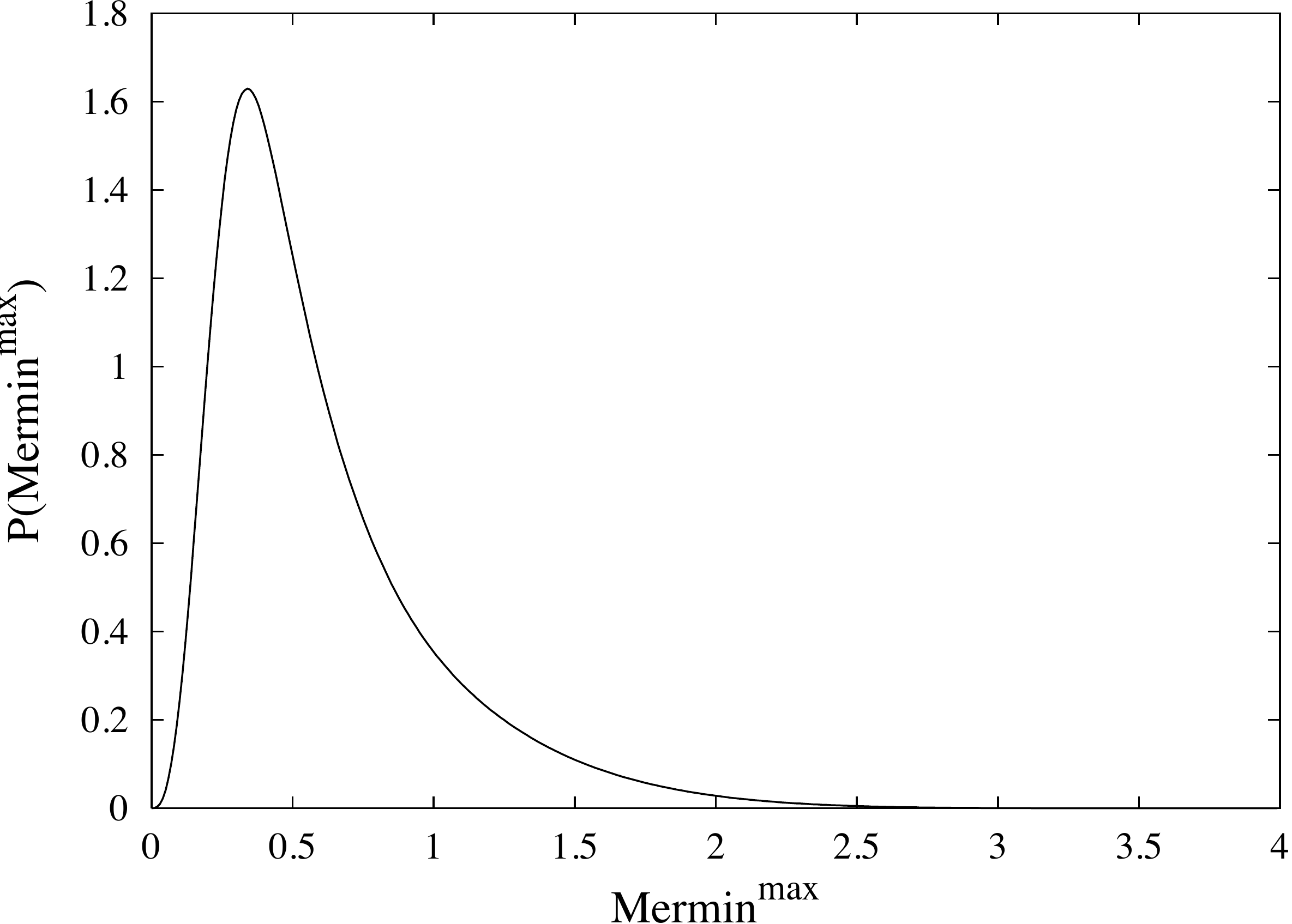}
\caption{Probability (density) distribution for the bound to nonlocality measure $Mermin^{\max}$ (\ref{MerminMax}) 
for all pure and mixed states of three qubits diagonal in the Mermin basis. See text for details.}
\label{fig2}
\end{center}
\end{figure}

A complementary Monte Carlo numerical survey was performed 
over a set of $10^9$ sample states generated with a uniform distribution for the set $\{ \lambda_i\}$ of the concomitant eigenvalues. This exhaustive, 
random exploration confirms the previous result. Fig. 3 depicts the probability (density) distribution 
for nonlocality measure $Mermin^{\max}$ (\ref{MerminMax}). Notice the strong biased behavior towards no Mermin inequality violation, as well as the relative 
scarcity of those states with some nonlocality.     

A previous work \cite{coreans} considered too the connexion between distillability and violation of the Mermin inequality for three qubits. It was concluded there 
that for a particular 4 parameter three qubit states, nonlocality implied Mermin-distillability. With our analysis, which embraces a more general class of states, we find that 
this is not the case. In point of fact, our numerical exploration obtains a probability 0.293 to find distillable states with no violation of the Mermin inequality, whereas those states 
that being distillable and achieving some nonlocality nearly possess  a zero-measure (probability of 0.008). The vast majority of states are found, with probability 0.698,  
both bound entangled and with no violation of the Mermin inequality. 

In view of our results, it seems plausible to assume that mixed states of three qubits with high amounts of Mermin nonlocality, which are likely to be Mermin-diagonal, possess no 
bound entanglement and are thus non-distillable.

\section{FOUR QUBITS}


The first Bell inequality for four qubits was derived by Mermin, Ardehali, Belinskii and Klyshko \cite{MABK}. It constitutes of 
four parties with two dichotomic outcomes each, being maximum for the generalized GHZ state $(|0000\rangle + |1111\rangle)/\sqrt{2}$.
The Mermin-Ardehali-Belinskii-Klyshko (MABK) inequality reads as $Tr(\rho {\cal B}_{MABK}) \leq 4$, where ${\cal B}_{MABK}$ is the MABK operator

\begin{equation} \label{MABK}
\begin{split}
B_{1111}&-B_{1112} -B_{1121}-B_{1211}-B_{2111}-B_{1122}-B_{1212}\\
&-B_{2112}-B_{1221}-B_{2121}-B_{2211}+B_{2222}+B_{2221}\\
&+B_{2212}+B_{2122}+B_{1222},
\end{split}
\end{equation}

\noindent with $B_{uvwx} \equiv {\bf u} \cdot {\bf \sigma} \otimes {\bf v} \cdot {\bf \sigma} \otimes {\bf w} \cdot {\bf \sigma}\otimes {\bf x} \cdot {\bf \sigma}$ 
with ${\bf \sigma}=(\sigma_x,\sigma_y,\sigma_z)$ being the usual Pauli matrices. As in previous instances, we shall define the following quantity

\begin{equation} \label{MABKMax}
 MABK^{\max} \equiv \max_{\bf{a_j},\bf{b_j}}\,\,Tr (\rho {\cal B}_{MABK})
\end{equation}

\noindent as a measure for the nonlocality content for a given state $\rho$ of four qubits. ${\bf a_j}$ and ${\bf b_j}$ are unit vectors
in $\mathbb{R}^3$. MABK inequalities are such that they constitute extensions of previous inequalities with the requirement that generalized 
GHZ states must maximally violate them. New inequalities for four qubits have appeared recently (see Ref. \cite{MABKnew}) that possess some other states 
required for optimal violation. In the present study we limit our interest to the MABK inequality, although new ones could be incorporated in 
order to offer a broader perspective. However, with respect to entanglement, little is know for the quadripartite case, and thus little 
comparison can be done.

\subsection{Nonlocality for four qubit states. Extension to generalized GHZ states}

The maximization of the MABK inequality $Tr(\rho {\cal B}_{MABK}) \leq 4$ for four qubits mixed states is done along similar lines as 
previously performed for bipartite and tripartite cases. Demanding maximum amount of nonlocality (\ref{MABKMax}) is tantamount as 
computing their optimum values for those states that concentrate nonlocal correlations. In the case of four qubits, 
those maximally correlated states are the ones which are diagonal in the Bell$_4$ basis defined by 
$|\Psi_j^{ \pm } \rangle =(|j\rangle \pm |15-j\rangle)/\sqrt{2}$. Therefore we shall consider the following class 
of four qubits mixed states

\begin{equation} \label{rhodiag4}
\rho_{MABK}^{(diag)}= \sum_{j=0}^7 (\lambda_j^{+}|\Psi_j^{+} \rangle \langle \Psi_j^{+}| + \lambda_j^{-}|\Psi_j^{-} \rangle \langle \Psi_j^{+}|),
\end{equation}
 
 \noindent with ordered eigenvalues $\lambda_{i+1} \geq \lambda_{i}$.

Computation of one term $B_{\alpha \beta \gamma \delta} \equiv {\bf \alpha} \cdot {\bf \sigma} \otimes {\bf \beta} \cdot \bf{\sigma} 
\otimes {\bf \gamma} \cdot \bf{\sigma} \otimes {\bf \delta} \cdot \bf{\sigma}$ of the MABK operator (\ref{MABK}) for the Bell$_4$ basis reads as

\begin{equation}\label{Qs}
\begin{split}
\langle B_{\alpha \beta \gamma \delta}  \rangle_0^{\pm}&= \alpha_z\beta_z\gamma_z\delta_z \pm Re[\alpha^{+} \beta^{+} \gamma^{+} \delta^{+}],\\
\langle B_{\alpha \beta \gamma \delta}  \rangle_1^{\pm}&= -\alpha_z\beta_z\gamma_z\delta_z \pm Re[\alpha^{+} \beta^{+} \gamma^{+} \delta^{-}],\\
\langle B_{\alpha \beta \gamma \delta}  \rangle_2^{\pm}&= -\alpha_z\beta_z\gamma_z\delta_z \pm Re[\alpha^{+} \beta^{+} \gamma^{-} \delta^{+}],\\ 
\langle B_{\alpha \beta \gamma \delta}  \rangle_3^{\pm}&=  \alpha_z\beta_z\gamma_z\delta_z \pm Re[\alpha^{+} \beta^{+} \gamma^{-} \delta^{-}],\\
\langle B_{\alpha \beta \gamma \delta}  \rangle_4^{\pm}&= -\alpha_z\beta_z\gamma_z\delta_z \pm Re[\alpha^{-} \beta^{+} \gamma^{-} \delta^{-}],\\ 
\langle B_{\alpha \beta \gamma \delta}  \rangle_5^{\pm}&=  \alpha_z\beta_z\gamma_z\delta_z \pm Re[\alpha^{+} \beta^{-} \gamma^{+} \delta^{-}],\\
\langle B_{\alpha \beta \gamma \delta}  \rangle_6^{\pm}&=  \alpha_z\beta_z\gamma_z\delta_z \pm Re[\alpha^{-} \beta^{+} \gamma^{+} \delta^{-}],\\ 
\langle B_{\alpha \beta \gamma \delta}  \rangle_7^{\pm}&= -\alpha_z\beta_z\gamma_z\delta_z \pm Re[\alpha^{-} \beta^{+} \gamma^{+} \delta^{+}],
\end{split}
\end{equation}

Gathering all pure-state expectation values, $Tr(\rho_{MABK}^{(diag)} {\cal B}_{MABK})$ is of the form 

\begin{equation}\label{Qs}
\sum_{j=0}^{7} \pm f(\Pi_j^z) (\lambda_j^{+} + \lambda_j^{-}) + g({\bf \alpha,\beta,\gamma,\delta}) (\lambda_j^{+} + \lambda_j^{-}),
\end{equation}

\noindent with $f(\cdot)$ is a real function of the product of all z-components of the four parties' settings, and $g(\cdot)$ represents a real function of several 
products of all parties' components, all of them according to the special form of the MABK Bell inequality operator (\ref{MABK}). 

The maximum value of (\ref{Qs}) is attained with $f(\Pi_j^z)=0$, that is, no z-dependency. Similarly to the calculations carried out in 
Appendices I and II, the optimum value of the violation of the MABK inequality $Tr(\rho {\cal B}_{MABK}) \leq 4$ for mixed states (\ref{rhodiag4}) is of the form

\begin{equation}\label{4qubits}
\max_{\bf{a_j},\bf{b_j}}\,\,Tr (\rho {\cal B}_{MABK}) \leqq 4\sqrt{2} \sqrt{\sum_{j=0}^7 (\lambda_j^{+} -\lambda_j^{-})^2}.
\end{equation}

\noindent The exact form for (\ref{4qubits}), obtained by recourse to convex optimization, is extremely complicated 
(combination of rational functions of radicals involving integer powers of $(\lambda_j^{+} - \lambda_j^{-})$). 
As in the case of three qubits, the previous bound is an excellent one, and hence we can consider the equality 
in (\ref{4qubits}) as a close or exact measure of the amount of nonlocality present in a Bell$_4$ diagonal mixed state 
of four qubits. 

When comparing this last result for four qubits with those of two and three qubits, we see that all three cases involve the same functional 
form for the eigenvalues of the mixed multipartite state diagonal in the concomitant maximally correlated basis. This is not surprising 
since the Bell inequalities considered so far are multipartite generalizations of the CHSH Bell inequality \cite{MABK}. Therefore we can conjecture the 
form for the maximal violation of the n-party generalized MABK inequality $Tr(\rho {\cal B}_{MABK}) \leq 4$ for a mixed state diagonal 
in the corresponding maximally correlated basis.\newline

\noindent {\bf Conjecture.} {\it The maximum amount of violation of  n-party generalized MABK inequalities for diagonal n-qubit mixed states is equal or less to} 
\begin{equation}\label{Nqubits}
 2^{\frac{n+1}{2}} \sqrt{\sum_{j=0}^{2^{n-1}-1} (\lambda_j^{+} -\lambda_j^{-})^2}.
\end{equation}
\newline 

Generalized GHZ states are those states of N parties 

\begin{equation}\label{GHZN}
\sqrt{p}|0\rangle + \sqrt{1-p} |2^{N-1}-1\rangle = \cos \alpha|0\rangle + \sin \alpha |2^{N-1}-1\rangle
\end{equation}

\noindent who maximally violate the MABK inequality \cite{MABK} $Tr(\rho {\cal B}_{MABK_N}) \leq {\cal B}_{MABK_N}^{LVM}$, where 
${\cal B}_{MABK_N}^{LVM}$ stands for the maximum violation allowed by a local variable model. 
We denote the quantal maximum violation \cite{MABK} by ${\cal B}_{MABK_N}^{QM}=2^{\frac{N+1}{2}}$.

The maximum value for the corresponding MABK inequality for states (\ref{GHZN}) is obtained by exactly following the optimization procedures 
carried out in the Appendices. In point of fact, GHZ states (\ref{GHZN}) are linear combinations of two pure states that can be written in 
the corresponding maximally correlated basis for that particular number of parties. That is, we can rewrite for convenience states (\ref{GHZN}) in the form 

\begin{equation}\label{GHZNbis}
\begin{split}
&\bigg( \sqrt{\frac{p}{2}}+\sqrt{\frac{1-p}{2}} \bigg) |\Phi_0^{+}\rangle_N +  \bigg( \sqrt{\frac{p}{2}}-\sqrt{\frac{1-p}{2}} \bigg) |\Phi_0^{-}\rangle_N\\
&=\sqrt{\lambda_1}|\Phi_0^{+}\rangle_N + \sqrt{\lambda_2} |\Phi_0^{-}\rangle_N, 
\end{split}
\end{equation}

\noindent with $\lambda_{1,2}=\frac{1}{2} \pm \sqrt{p(1-p)}$ ($\lambda_1 \geq \lambda_2$). This new form 
enables us to treat generalized GHZ as linear combinations of maximally correlated states in a similar fashion as performed 
for two qubit states. 

After some algebra, we obtain that the leading term in the violation goes as $2\,{\cal B}_{MABK_N}^{QM}\sqrt{p(1-p)}$ 
(symmetric around $p=\frac{1}{2}$), from which we reobtain, after equating it 
to ${\cal B}_{MABK_N}^{LVM}$, the well known result \cite{Scarani} $\sin 2\alpha \leq 1/\sqrt{2^{N-1}}$.
Thus, by employing our optimization procedure, we not only recover the range where 
generalized GHZ states violate a Bell inequality, but also obtain its exact amount.\newline

As far as entanglement for four qubits is concerned, we encounter a considerable discrepancy between maximum entanglement and nonlocality. 
First of all, it is well known that no proper entanglement measure is operational yet for states living in arbitrary Hilbert spaces. 
However, some measures (for pure states) based on partitions of the system have been advanced, such as the
so-called global entanglement (GE), which describes the average entanglement of each qubit of the system with
the remaining $N-1$ qubits. The GE measure is widely
regarded as a legitimate $N$-qubit entanglement measure \cite{Brennen,Weistein1,Weistein2}. Having a general mixed state implies that the previous 
measures do not apply as such. To overcome this fact we require the extension of these partition-based entanglement measures by
recourse to the usual convex roof \cite{Nielsen} defined over some given set of pure states. Given the extraordinary numerical effort 
that this procedure would imply, an alternative measure is given by the sum of the von Neumann entropy
of the reduced density matrices of all individual qubits, that is, $S_{vN}=\sum_i^N S_{vN}(\rho_i)$. Incidentally, it
has also been considered as a proper entanglement measure in the literature \cite{final1}.

It is in this sense that, when employing the sum of the von Neumann entropy of all partitions of a state, 
the maximum is reached \cite{HS} for a particular (pure) state different from the generalized GHZ state for four qubits 
(which is not the case for two or three qubits). Therefore we encounter that maximum entanglement does not correspond to 
maximum nonlocality for four qubits already at the level of pure states. The analysis for mixed states --not performed here-- 
would simply confirm this result.


\section{CONCLUSIONS}

In the present work we have studied how nonlocality is present in systems of two, three and four qubits. We have 
exhaustively explored several aspects that are shared by quantum entanglement and 
nonlocality --measured by the maximum violation of a Bell inequality-- as well as pointed out 
those ones that differentiate these two magnitudes. By highlighting those issues that concern 
entanglement and nonlocality, we shed a new light on the connections that exists between these two 
concepts that play a paramount role in quantum-information theory and, in turn, in the foundations 
of quantum mechanics. 

By means of a new optimization method, we have computed the maximum violation of the CHSH inequality for two qubit 
systems and obtained the concomitant maximal states MNMS within the context defined by the degree of mixture, as measured by 
either the participation ration $R$ of the maximum eigenvalue $\lambda_{\max}$ of the state $\rho$. The direct comparison with 
MEMS states illustrates an anomaly that appears between entanglement and nonlocality already for mixed states of two qubits, 
enhanced by the information-theoretic tasks limitations that appear in the study of bipartite states in the infinite $XY$ model. The study of 
nonlocality for linear combinations of pure states of two qubits allowed us to compare how both nonlocality and entanglement are 
distributed in the pure state superposition. 

The extension to three qubit states was done along similar lines, employing the maximal violation of the Mermin inequality as a nonlocality 
measure. Analogous computations allowed us to obtain the expression for the nonlocality present in mixed states diagonal in the GHZ basis. 
We obtained that the generalized Werner states for three qubits possess maximum nonlocality for a given value of the degree of mixture, 
contrary to the two qubit scenario. Also, we extend a previous result concerning distillability and nonlocality in the light of quantum entanglement.

The study of four qubit systems was performed following the same steps as in the previous two cases: by maximizing the MABK inequality for four qubits, 
we obtained the maximum violation of this nonlocality measure for mixed states diagonal in the Bell$_4$ basis. As a consequence, a careful quantitative 
analysis is performed for generalized GHZ states as well. As far as entanglement is concerned, we observe the first discrepancy between maximum 
entangled and optimal nonlocality already for pure sates of four qubits. Obviously, the MABK inequalities are not the only existing Bell inequalities for
states of arbitrary number of qubits, but because it constitutes a simple generalization
of the CHSH inequality (nevertheless, some authors introduce other inequalities 
that incorporate the MABK ones as special cases \cite{RefZukowskiGHZnXinos}), 
it has been enough to make use of this particular family of inequalities to illustrate the basic 
results of the present work. Some further work will be required regarding different multipartite 
Bell inequalities.

Despite the fact that for small quantum systems we recognize a simple correlation between entanglement and nonlocality, 
the entire situation becomes more involved when the dimension of the Hilbert space of the system or subsystems augments. 
This fact is certainly transcendental for several information-theoretic task require the presence of either quantities. Physical 
situations such as the one encountered in the $XY$ model, were null nonlocality for two or three parties is compatible 
with non-zero entanglement, do not contribute to unify the ultimate quantum correlations that define the state of a 
quantum system. Rather, we are tempted to regard nonlocality and entanglement as different quantum resources in view of the 
undefined limits between them.

On the whole, however, many aspects that also concern nonlocality and entanglement have not been considered here. One 
such example could be the so called {\it monogamy of entanglement}, a fundamental property stating that 
if two quantum systems are maximally correlated (maximum entanglement), then they cannot be correlated 
with a third party. This is, for instance, the basis for secure quantum key distribution based on 
entanglement \cite{E91}. The fact that this trade-off also occurs for nonlocality \cite{TonerBell} 
in the multipartite case constitutes an issue that surely deserves future study.

\section*{Acknowledgements}

J. Batle acknowledges fruitful discussions with J. Rossell\'{o} and M. del M. Batle. M. Casas acknowledges 
partial support under project FIS2008-00781/FIS (MICINN) and FEDER (EU).

\section*{APPENDIX I}

The goal of this appendix is to derive the maximum violation of the CHSH inequality (\ref{CHSH_LVM}) for two qubit systems. Such endeavor 
might render somewhat difficult the study of the general instance, but this is not the case for there is no need to explore the whole space of mixed states of 
two qubits. Since we require $Tr(\rho{\cal B}_{CHSH})$, which is a convex function of the two qubit state $\rho$, to be maximum, 
it suffices to consider those states that concentrate all quantum correlations after the action of a depolarizing channel \cite{depol}. 
This class of states are, as expected, the Bell diagonal states.

The optimization is taken over the two observers' settings $\{{\bf a_j},{\bf b_j}\}$, which are real unit vectors in $\mathbb{R}^3$. We choose them to be 
of the form $(\sin\theta_k \cos\phi_k,\sin\theta_k \sin\phi_k,\cos\theta_k)$. With this parameterization, the problem consists in finding 
the supremum of $Tr(\rho{\cal B}_{CHSH})$ over the $\{k=1\dotsm 8\}$ angles of $\{  {\bf a_1},{\bf b_1},{\bf a_2},{\bf b_2} \}$ 
that appear in (\ref{CHSH_QM}).

The general form entering the Bell operator (\ref{CHSH_QM}) for one single entry is 
of the kind ${\bf \alpha} \cdot {\bf \sigma} \otimes {\bf \beta} \cdot \bf{\sigma}$. 
Written in the computational basis $\{ |0\rangle, |1\rangle\}$, we have

\begin{equation} \label{dos1}
\left( \begin{array}{cccc}
\alpha_z\beta_z & \alpha_z\beta^{-} & \alpha^{-}\beta_z & \alpha^{-}\beta^{-}\\
\alpha_z\beta^{+} & -\alpha_z\beta_z & \alpha^{-}\beta^{+} & -\alpha^{-}\beta_z\\
\alpha^{+}\beta_z & \alpha^{+}\beta^{-} & -\alpha_z\beta_z & -\alpha_z\beta^{-}\\
\alpha^{+}\beta^{+} & -\alpha^{+}\beta_z & -\alpha_z\beta^{+} & \alpha_z\beta_z 
\end{array} \right),
\end{equation} 

\noindent with $\alpha^{\pm}=\alpha_x \pm i\alpha_y$ and $\beta^{\pm}=\beta_x \pm i\beta_y$. The evaluation of (\ref{dos1}) for all states in the 
Bell basis reads as

\begin{equation}\label{dos2}
\begin{split}
\langle {\bf \alpha} \cdot {\bf \sigma} \otimes {\bf \beta} \cdot \bf{\sigma}  \rangle_{\Phi^{\pm}} &= \alpha_z\beta_z \pm Re[\alpha^{+} \beta^{+}], \\
\langle {\bf \alpha} \cdot {\bf \sigma} \otimes {\bf \beta} \cdot \bf{\sigma}  \rangle_{\Psi^{\pm}} &= -\alpha_z\beta_z \pm Re[\alpha^{+} \beta^{-}].
\end{split}
\end{equation}

The expression for $Tr(\rho_{Bell}^{(diag)} {\cal B}_{CHSH})$, with eigenvalues $\lambda_1 \geq \lambda_2 \geq \lambda_3 \geq \lambda_4$, can be 
cast as 

\begin{equation}\label{dos3}
\begin{split}
  & \big( (\lambda_1 - \lambda_4)-( \lambda_2 - \lambda_3) \big)   [a_1^x (b_1^x+b_2^x) + a_2^x (b_1^x - b_2^x)]\\
+& \big( (\lambda_2 + \lambda_3)-( \lambda_1 + \lambda_4) \big) [a_1^y (b_1^y+b_2^y) + a_2^y (b_1^y - b_2^y)]\\
+& \big( (\lambda_1 - \lambda_4)+( \lambda_2 - \lambda_3) \big)  [a_1^z (b_1^z+b_2^z) + a_2^z (b_1^z - b_2^z)].
\end{split}
\end{equation}

\noindent Eigenvalue coefficients in the first and second terms of (\ref{dos3}) are strictly positive, whereas in the second one it is undefined. By rearranging 
terms in (\ref{dos3}), we obtain

\begin{equation} \label{dos4}
\begin{split}
   (\lambda_1 - \lambda_4) & \big[a_1^z (b_1^z+b_2^z) + a_2^z (b_1^z - b_2^z) \\
                                               +& \big( a_1^x (b_1^x+b_2^x) + a_2^x (b_1^x - b_2^x \big) \big]\\
                                               &\\
+ (\lambda_2 - \lambda_3) &  \big[a_1^z (b_1^z+b_2^z) + a_2^z (b_1^z - b_2^z)\\ 
                                               -& \big( a_1^x (b_1^x+b_2^x) + a_2^x (b_1^x - b_2^x) \big)\big]\\
                                               &\\
+   \Delta_\lambda & \big[a_1^y (b_1^y+b_2^y) + a_2^y (b_1^y - b_2^y)\big], 
\end{split}
\end{equation}

\noindent with $\Delta_\lambda \equiv (\lambda_2 + \lambda_3)-( \lambda_1 + \lambda_4)$ possessing no clear sign, which would imply some further 
insight. However, on the contrary, this fact points out that no y-dependency makes (\ref{dos4}) even greater. Also, the symmetry in (\ref{dos4}) allows us to choose 
the alignment of one of the settings. From inspection of (\ref{dos4}), we therefore optimize it by choosing 
$\{\a_1=(1,0,0), \a_2=(0,0,-1), \b_1=(b_1^x,0,-b_2^z), \b_2=(b_2^x,0,b_2^z) \}$. 

The final concomitant result amounts to 

\begin{equation} \label{dos5}
\begin{split}
 \max_{\bf{a_j},\bf{b_j}}&\,\,Tr (\rho {\cal B}_{CHSH}) =\\
  \max_{b_1^x,b_2^z}& \,\,2(\lambda_1 - \lambda_4)[b_2^z+b_1^x] + 2(\lambda_2 - \lambda_3)[b_2^z-b_1^x] =\\
  \max_{\Theta}&\big[  2\sqrt{2}(\lambda_1 - \lambda_4) \cos(\Theta) + 2\sqrt{2}(\lambda_2 - \lambda_3) \sin(\Theta) \big]=\\
 & 2\sqrt{2} \sqrt{(\lambda_1 - \lambda_4)^2 \,\, + \,\,(\lambda_2 - \lambda_3)^2}.
 \end{split}
 \end{equation}
 \newline

 In the case where nonlocality is to be found in linear combinations of pure states of two qubits, we shall perform a similar analysis. Our starting point 
 is the matrix of expectation values of elements  ${\bf \alpha} \cdot {\bf \sigma} \otimes {\bf \beta} \cdot \bf{\sigma}$ (\ref{dos1}) in the Bell basis 
 $\{ |\Phi^{+}\rangle,|\Phi^{-}\rangle,|\Psi^{+}\rangle,|\Psi^{-}\rangle \}$


\begin{widetext}
\begin{equation}\label{dos6}
\left( \begin{array}{cccc}
\alpha_z\beta_z + Re[\alpha^{+} \beta^{+}] & i Im[\alpha^{+} \beta^{+}] & i Im[\alpha^{-}\beta_z + \alpha_z\beta^{-}] & Re[\alpha_z \beta^{+} - \alpha^{+} \beta_z]\\
-i Im[\alpha^{+} \beta^{+}] & \alpha_z\beta_z - Re[\alpha^{+} \beta^{+}] & Re[\alpha_z \beta^{+} + \alpha^{+} \beta_z] & i Im[ \alpha_z\beta^{-} + \alpha^{+}\beta_z ]\\
-i Im[\alpha^{-}\beta_z + \alpha_z\beta^{-}] & Re[\alpha_z \beta^{+} + \alpha^{+} \beta_z] & -\alpha_z\beta_z + Re[\alpha^{+} \beta^{-}] & i Im[\alpha^{-} \beta^{+}] \\
Re[\alpha_z \beta^{+} - \alpha^{+} \beta_z] & -i Im[ \alpha_z\beta^{-} + \alpha^{+}\beta_z ] & -i Im[\alpha^{-} \beta^{+}] & -\alpha_z\beta_z - Re[\alpha^{+} \beta^{-}]
\end{array} \right).
\end{equation}
\end{widetext}


Let us consider a general pure state of the form  

\begin{equation} \label{dos7}
|\phi\rangle = \sqrt{\lambda_1}|\Phi^{+}\rangle + \sqrt{\lambda_2}|\Phi^{-}\rangle 
+ \sqrt{\lambda_3}|\Psi^{+}\rangle + \sqrt{\lambda_4} |\Psi^{-}\rangle,    
\end{equation} 

\noindent with real coefficients $\{ \sqrt{\lambda_i} \}$ such that $\lambda_1+\lambda_2+\lambda_3+\lambda_2=1$. 
We will demand the latter to be sorted in decreasing value though, as we shall see, this is not mandatory. 
The general case with complex coefficients is somewhat more involved. However, for most practical purposes, 
it will suffice to consider real states of the form (\ref{dos7}).

The fact of having real coefficients in (\ref{dos7}) greatly simplifies the expectation value 
$\langle \phi|{\cal B}_{CHSH}|\phi \rangle$. Its general term term $\langle \phi| {\bf \alpha} \cdot {\bf \sigma} 
\otimes {\bf \beta} \cdot \bf{\sigma} |\phi \rangle$ (\ref{dos1}) is of the form
    
\begin{equation} \label{dos8}
\begin{split}
\lambda_1 & \big(\alpha_z\beta_z + Re[\alpha^{+} \beta^{+}] \big) + \lambda_2  \big(\alpha_z\beta_z - Re[\alpha^{+} \beta^{+}] \big) \\
\lambda_3 & \big(-\alpha_z\beta_z + Re[\alpha^{+} \beta^{-}] \big) + \lambda_4  \big(-\alpha_z\beta_z - Re[\alpha^{+} \beta^{-}] \big) \\
+& \sqrt{\lambda_2 \lambda_3} 2 Re[\alpha_z \beta^{+} + \alpha^{+} \beta_z] + \sqrt{\lambda_1 \lambda_4} 2 Re[\alpha_z \beta^{+} - \alpha^{+} \beta_z].
\end{split}
\end{equation}

\noindent Notice that, in view of (\ref{dos8}), that the optimization of the CHSH inequality for linear combinations of pure states in the 
Bell basis is almost identical to the corresponding mixed state case (differing in the last two terms of (\ref{dos8})).

Proceeding as before, optimization of $\langle \phi|{\cal B}_{CHSH}|\phi \rangle$ returns the value

   
\begin{widetext}
\begin{equation}\label{dos9}
2\sqrt{2} \sqrt{(\lambda_1 - \lambda_4)^2 + (\lambda_2 - \lambda_3)^2 +4[\sqrt{\lambda_1 \lambda_4}]^2+4[\sqrt{\lambda_2 \lambda_3}]^2}  
= 2\sqrt{2} \sqrt{(\lambda_1 + \lambda_4)^2 + (\lambda_2 + \lambda_3)^2}.
\end{equation}
\end{widetext}

\section*{APPENDIX II}

In this appendix we shall derive the explicit form for the maximum amount (\ref{MerminMax}) of violation of the Mermin inequality 
for a three qubit state. As expected, since Tr($\rho {\cal B}_{Mermin}$) is a convex function of the quantum state $\rho$, its maximum 
is obtained only for pure states, namely, the whole class of states forming the Mermin-basis $|\Psi_j^{ \pm } \rangle =(|j\rangle \pm |7-j\rangle)/\sqrt{2}$. 
In view of this observation, we shall consider instead what is the maximum violation attained for mixed states diagonal in this basis, that is, 
states of the class $\rho_{Mermin}^{(diag)}$ (\ref{rhodiag}). Another argument for studying these states is that any initial state $\rho$ can be converted into one 
in the class by means of LOCC.

Optimization of $Mermin^{\max}$ (\ref{MerminMax}) for states $\rho_{Mermin}^{(diag)}$ (\ref{rhodiag}) is carried out in the same fashion as 
in the previous bipartite case. Once the observers' settings $\{{\bf a_j},{\bf b_j}\}$, which are real unit vectors in $\mathbb{R}^3$, are parameterized in spherical coordinates 
$(\sin\theta_k \cos\phi_k,\sin\theta_k \sin\phi_k,\cos\theta_k)$, the problem consists in finding the supremum of (\ref{MerminMax}) over the set of $\{k=1\dotsm12\}$ 
possible angles for $\{  {\bf a_1},{\bf b_1},{\bf a_2},{\bf b_2},{\bf a_3},{\bf b_3} \}$ in (\ref{Mermin}). 

To start with, let us write a generic element of the Mermin operator (\ref{Mermin}) of the form 
$B_{\alpha \beta \gamma} \equiv {\bf \alpha} \cdot {\bf \sigma} \otimes {\bf \beta} \cdot \bf{\sigma} \otimes {\bf \gamma} \cdot {\bf \sigma}$ in 
the computational basis $\{ |0\rangle, |1\rangle\}$ defined by the $z$-projections of ${\bf \sigma}=(\sigma_x,\sigma_y,\sigma_z)$. 
$B_{\alpha \beta \gamma}$ reads as 

\begin{equation} \label{tres}
\left( \begin{array}{cccc}
\alpha_z\beta_z & \alpha_z\beta^{-} & \alpha^{-}\beta_z & \alpha^{-}\beta^{-}\\
\alpha_z\beta^{+} & -\alpha_z\beta_z & \alpha^{-}\beta^{+} & -\alpha^{-}\beta_z\\
\alpha^{+}\beta_z & \alpha^{+}\beta^{-} & -\alpha_z\beta_z & -\alpha_z\beta^{-}\\
\alpha^{+}\beta^{+} & -\alpha^{+}\beta_z & -\alpha_z\beta^{+} & \alpha_z\beta_z 
\end{array} \right) 
\otimes \left(\begin{array}{cc}
\gamma_z & \gamma^{-} \\
\gamma^{+} &  -\gamma_z 
\end{array} \right), 
\end{equation} 

\noindent with $\alpha^{\pm}=\alpha_x \pm i\alpha_y$, $\beta^{\pm}=\beta_x \pm i\beta_y$, and $\gamma^{\pm}=\gamma_x \pm i\gamma_y$ 
being ``rising'' and ``lowering'' terms in the x-y plane. Nonlocality measure (\ref{MerminMax}) for diagonal states (\ref{rhodiag}) is computed 
by recourse to four $B_{\alpha \beta \gamma}$s in (\ref{tres}) for different configurations of vectors. 

The evaluation of (\ref{MerminMax}) for diagonal states (\ref{rhodiag}) amounts to compute the expectation value 
$\langle \Psi_j^{\pm}| B_{\alpha \beta \gamma} |\Psi_j^{\pm}\rangle \equiv \langle \alpha \beta \gamma  \rangle_j^{\pm}$ for all states $\{ |\Psi_j^{\pm}\rangle \}$ 
in the Mermin basis and several vector configurations. The positions in $ \langle \alpha \beta \gamma  \rangle_j^{\pm}$ 
are such that $\alpha$, $\beta$ and $\gamma$ correspond to the first, second and third observer, respectively. 
This computation returns

\begin{equation}\label{Mermin1}
\begin{split}
\langle \alpha \beta \gamma  \rangle_0^{\pm}&= \pm Re[\alpha^{+} \beta^{+} \gamma^{+}], \, \langle \alpha \beta \gamma  \rangle_1^{\pm}= \pm Re[\alpha^{+} \beta^{+} \gamma^{-}]\\
\langle \alpha \beta \gamma  \rangle_2^{\pm}&= \pm Re[\alpha^{+} \beta^{-} \gamma^{+}], \, \langle \alpha \beta \gamma  \rangle_3^{\pm}= \pm Re[\alpha^{-} \beta^{+} \gamma^{+}].
\end{split}
\end{equation}

\noindent As we can observe, the observers settings can be two-dimensional (he have no z-dependency).  

From previous definitions, we now write $Tr(\rho_{Mermin}^{(diag)} {\cal B}_{Mermin})$ as
\begin{equation}\label{Mermin2}
\begin{split}
&(\lambda_0^{+}-\lambda_0^{-}) \big[ \langle \a \a \a  \rangle_0^{+} - \langle \a \b \b  \rangle_0^{+} - \langle \b \a \b  \rangle_0^{+} - \langle \b \b \a  \rangle_0^{+}\big]\\
+&(\lambda_1^{+}-\lambda_1^{-}) \big[ \langle \a \a \a  \rangle_1^{+} - \langle \a \b \b  \rangle_1^{+} - \langle \b \a \b  \rangle_1^{+} - \langle \b \b \a  \rangle_1^{+}\big]\\
+&(\lambda_2^{+}-\lambda_2^{-}) \big[ \langle \a \a \a  \rangle_2^{+} - \langle \a \b \b  \rangle_2^{+} - \langle \b \a \b  \rangle_2^{+} - \langle \b \b \a  \rangle_2^{+}\big]\\
+&(\lambda_3^{+}-\lambda_3^{-}) \big[ \langle \a \a \a  \rangle_3^{+} - \langle \a \b \b  \rangle_3^{+} - \langle \b \a \b  \rangle_3^{+} - \langle \b \b \a  \rangle_3^{+}\big].
\end{split}
\end{equation}

\noindent The explicit evaluation of the previous quantity can be cast as

\begin{equation}\label{Mermin3}
\begin{split}
(\lambda_{0/1}^{+}-\lambda_{0/1}^{-}) \big[ 
 & [(a_1^x a_2^x - a_1^y a_2^y) a_3^x \mp (a_1^x a_2^y + a_1^y a_2^x) a_3^y] \\
-& [(a_1^x b_2^x - a_1^y b_2^y) b_3^x \mp (a_1^x b_2^y + a_1^y b_2^x) b_3^y] \\
-& [(b_1^x a_2^x - b_1^y a_2^y) b_3^x \mp (b_1^x a_2^y + b_1^y a_2^x) b_3^y] \\
-& [(b_1^x b_2^x - b_1^y b_2^y) a_3^x \mp (b_1^x b_2^y + b_1^y b_2^x) a_3^y] \big] \\
+&\\
(\lambda_{2/3}^{+}-\lambda_{2/3}^{-}) \big[ 
 & [(a_1^x a_2^x + a_1^y a_2^y) a_3^x \pm (a_1^x a_2^y - a_1^y a_2^x) a_3^y] \\
-& [(a_1^x b_2^x + a_1^y b_2^y) b_3^x \pm (a_1^x b_2^y - a_1^y b_2^x) b_3^y] \\
-& [(b_1^x a_2^x + b_1^y a_2^y) b_3^x \pm (b_1^x a_2^y - b_1^y a_2^x) b_3^y] \\
-& [(b_1^x b_2^x + b_1^y b_2^y) a_3^x \pm (b_1^x b_2^y - b_1^y b_2^x) a_3^y] \big].
\end{split}
\end{equation}

\noindent Since the Mermin inequality settings are such that it must posses rotationally 
invariance (x-y plane), we are free to fix one of them. In view of (\ref{Mermin3}), we choose 
$\a_2=(-1,0,0)$. Also, differences in each term of (\ref{Mermin3}) must be maximum in absolute value, 
which is compatible with fixing $\b_2=(0,1,0)$. Further calculations imply a configuration of the type 
$\{\a_1=(a_1^x,a_1^y,0), \a_3=(a_3^x,a_3^y,0), \b_1=(-a_1^x,a_1^y,0), \b_3=(a_3^x,-a_3^y,0) \}$. 
Expected value $Tr(\rho_{Mermin}^{(diag)} {\cal B}_{Mermin})$ greatly simplifies from (\ref{Mermin3}) 
into

\begin{equation}\label{Mermin4}
\begin{split}
(\lambda_{0/1}^{+}-\lambda_{0/1}^{-}) 
 & [2 | a_1^x a_3^x | + 2 | a_1^y a_3^x | \pm 2 | a_1^x a_3^y | \pm 2 | a_1^y a_3^y |] \\
+&\\
(\lambda_{2/3}^{+}-\lambda_{2/3}^{-}) 
& [2 | a_1^x a_3^x | - 2 | a_1^y a_3^x |  \mp 2 | a_1^x a_3^y| \pm 2 | a_1^y a_3^y | ].
\end{split}
\end{equation}

\noindent Notice that we have reduced our optimization problem to one which entails only two real quantities. 
By introducing explicit angles ($\phi,\psi$), and after some algebra, we obtain

\begin{equation} \label{Mermin5}
\begin{split}
 \max_{\bf{a_j},\bf{b_j}}\,\,Tr (\rho {\cal B}_{Mermin}) = 
 \max_{\phi,\psi} \,\,\,
 &(\lambda_0^{+}-\lambda_0^{-})  4 \sin \phi  \sin \psi \\
 +&(\lambda_1^{+}-\lambda_1^{-}) 4 \sin \phi  \cos \psi \\
 +&(\lambda_2^{+}-\lambda_2^{-}) 4 \cos \phi  \cos \psi \\
 +&(\lambda_3^{+}-\lambda_3^{-}) 4 \cos \phi  \sin \psi. 
 \end{split}
 \end{equation}

\noindent The solution of (\ref{Mermin5}) is obtained by recourse to the use of convex optimization techniques 
\cite{convex}. We do not worry about the signs in each term of (\ref{Mermin3}) since we have chosen, without loss of 
generality, the eigenvalues $\{ \lambda_j^{\pm} \}$ to be sorted in decreasing value. By solving the set of equations

\begin{equation} \label{Mermin6}
\begin{split}
\tan \phi &= \bigg(   \frac{4(\lambda_0^{+}-\lambda_0^{-}) \tan \psi + 4(\lambda_1^{+}-\lambda_1^{-})}{4(\lambda_2^{+}-\lambda_2^{-}) + 4(\lambda_3^{+}-\lambda_3^{-}) \tan \psi}    \bigg)\\
\tan \psi &= \bigg(   \frac{4(\lambda_0^{+}-\lambda_0^{-}) \tan \phi + 4(\lambda_3^{+}-\lambda_3^{-})}{4(\lambda_2^{+}-\lambda_2^{-}) + 4(\lambda_1^{+}-\lambda_1^{-}) \tan \phi}    \bigg) 
\end{split}
\end{equation}

\noindent we finally obtain the desired evaluation of (\ref{MerminMax}) for diagonal states (\ref{rhodiag}). Though the final result is 
rather cumbersome, we nevertheless derive an excellent bound. In view of the coefficients in (\ref{Mermin5}) (the sum of their 
squared values equals one), we provide the final result (\ref{MerminMax2}).


\begin{thebibliography}{0}
\bibitem {explicacio} It can be shown by induction that ${\cal B}_{CHSH}^{2n+1}=4^n  {\cal B}_{CHSH}$ and 
${\cal B}_{CHSH}^{2n}=4^n I_4$, where $I_4$ is the $4\times4$ identity matrix.    
\end{thebibliography}
\end{document}